\documentclass[trackchanges,twocolumn]{aastex701}

\usepackage{booktabs}
\usepackage{xcolor}

\begin{document}

\title{Cavity-driven Gas Transport in Cool-Core Clusters: Implications for Multiphase Gas Cooling}

\author[orcid=0000-0001-8176-7665,sname='Tamhane']{Prathamesh Tamhane}
\affiliation{Department of Physics and Astronomy, University of Alabama in Huntsville, 301 Sparkman Drive, Huntsville, AL 35899, USA}
\email[show]{pdt0003@uah.edu}  

\author[gname=Brian, sname='McNamara']{Brian McNamara} 
\affiliation{Waterloo center for Astrophysics, University of Waterloo, Waterloo, ON, N2L 3G1, Canada}
\affiliation{Department of Physics and Astronomy, University of Waterloo, Waterloo, ON, N2L 3G1, Canada}
\email{mcnamara@uwaterloo.ca}

\author[gname=Paul,sname=Nulsen]{Paul Nulsen}
\affiliation{ICRAR, Perth, Western Australia, 6009, Australia}
\affiliation{Harvard-Smithsonian center for Astrophysics, 60 Garden Street, Cambridge, MA 02138, USA}
\email{paulnulsen@gmail.com}


\begin{abstract}

We present a study of gas uplift by cavities and its role in promoting the formation of molecular gas in cool-core galaxy clusters. Using a sample of 24 clusters, we quantify the parameter space of gas uplift, including lifting masses, velocities, radii, and cooling times, and compare the displaced hot gas to observed molecular gas masses. Almost all clusters contain sufficient low-entropy gas within the central $\sim$10 kpc to account for the observed molecular gas. The molecular gas mass scales approximately linearly with the maximum gas mass displaced by cavities. Gas with cooling times of a few $10^8$ yr can be transported to typical H$\alpha$ filament radii of 10--30 kpc at characteristic velocities of 150--400 km s$^{-1}$, whereas gas with $t_{\rm cool}\sim10^9$ yr generally requires uplift to $\gtrsim60$ kpc. The mass expected to cool over the lifetimes of the currently observed cavities is comparable to the molecular reservoir in several systems, assuming that 50\% of the displaced gas mass is lifted and cools, whereas other systems contain excess molecular gas. This suggests that the molecular gas reservoirs may have accumulated over multiple AGN feedback cycles involving repeated gas uplift and cooling, or through additional processes. H$\alpha$ and infrared luminosities increase linearly with X-ray cooling luminosity, linking the multiphase emission to the cooling atmosphere. Overall, our results support a picture in which repeated AGN outbursts with gas uplift promote the condensation of low-entropy gas and help build molecular reservoirs in cool-core BCGs.

\end{abstract}



\section{Introduction} 

The centers of galaxy clusters host some of the most powerful active galactic nuclei (AGN) in the local Universe. Energetic feedback from the AGN is important for regulating the growth of massive galaxies \citep{best05,fabian12,mcnamara07,mcnamara12}. The effect of AGN can be seen as atomic and molecular galactic outflows observed in nearby galaxies \citep{feruglio10,cicone14,fiore17} as well as in the Brightest Cluster Galaxies (BCGs) residing in the centers of cool-core galaxy clusters \citep{russell19,olivares19,tamhane22}. While atomic and molecular outflows observed in nearby galaxies are driven by the hot wind and radiation pressure of the AGN \citep{fluetsch19}, the molecular flows observed in the Brightest Cluster Galaxies (BCGs) are primarily driven by the mechanical feedback of the AGN in the form of buoyantly rising X-ray cavities in the atmospheres of galaxy clusters.

The BCGs are embedded in atmospheres containing hot 10$^{7-8}$ K gas. This gas is expected to cool and form a large amount of molecular gas. However, the observed levels of molecular gas ($\sim$ 10$^{10}$ M$_\odot$) revealed by Atacama Large Millimeter/submillimeter Array (ALMA) \citep[see][for summary]{russell19} and IRAM \citep{edge01,salome03} are far below the theoretically predicted limits from pure cooling \citep{edge03}. Observations have since shown that the radio-mechanical feedback in the form of radio jets produced by the central AGN heats the surrounding gas through weak shocks, sound waves, turbulence and cosmic rays and balances cooling \citep{mcnamara07,mcnamara12,fabian12}. Nevertheless, some cooling persists giving rise to the formation of molecular gas. This molecular gas is essential for fueling star formation as well as the central AGN and serves as a link connecting these processes with the atmosphere from which it cooled \citep{mathews03,pizzolato05}. 

A substantial fraction of the ionized and molecular gas in BCGs lies in filaments that trail behind buoyantly rising bubbles formed by radio jets within the intra-cluster medium (ICM) of cool-core clusters \citep{russell19,olivares23}. These bubbles are visible as surface brightness depressions or cavities in X-ray images \citep{birzan04}. Recent observations showed that cool-core clusters shine in H$\alpha$ and CO emission when the cooling time of the hot gas in central regions falls below 5$\times$10$^8$ yr or central entropy falls below 30 keV cm$^2$ \citep{cavagnolo08,pulido18}. Initial models of multiphase gas formation in hot atmospheres proposed that cold gas condenses through local thermal instabilities, with condensation occurring when the ratio of cooling time to free-fall time falls below a critical threshold of $\sim10$ \citep{mccourt12,prasad15,voit15}. Subsequent observations, however, revealed substantial scatter in the observed $t_{\rm cool}/t_{\rm ff}$ ratios of systems hosting H$\alpha$ and CO emission, suggesting that condensation is not governed by a single universal threshold and that additional processes such as uplift, turbulence, and bulk gas motions likely play an important role in promoting cooling \citep[e.g.,][]{li14,valentini15,gaspari17,voit17}. Cooling time and entropy thresholds have therefore emerged as useful empirical predictors of the presence of multiphase gas.

However, \citet{martz20} showed that central atmospheres of Abell 2029, Abell 2151, RBS0540 and RBS0533 have short central cooling times and low central entropies, yet these systems are devoid of nebular emission or molecular gas. The X-ray atmospheres of those BCGs are fairly relaxed, suggesting that some form of atmospheric disturbance may be required to trigger or enhance gas condensation.

Cavities blown by radio jets are observed in almost all cool-core clusters and may play an important role in promoting gas cooling. Early simulations of AGN feedback showed that gas in the central regions of galaxies can be lifted by cavities to larger radii where AGN heating is less effective, allowing it to cool radiatively \citep{revaz08}. Building on this idea, \citet{mcnamara16} proposed a stimulated feedback mechanism in which cavities lift low-entropy gas to radii where the ratio of the cooling time to the infall time becomes less than unity ($t_{\rm cool}/t_{\rm I} \lesssim 1$), triggering thermal instabilities and condensation of cold gas. This mechanism is motivated by observations of cold gas filaments around or oriented toward X-ray cavities and naturally explains the filamentary morphology commonly seen in molecular gas in cool-core BCGs, as well as the observed correlation between the kinematics of the hot and warm ionized phases \citep{hitomi16,olivares25}. In this framework, the ability of radio jets to lift low-entropy gas may also explain why molecular gas flows in BCGs are substantially larger and more massive than those observed in other active galaxies \citep{tamhane22}, highlighting the importance of radio-mechanical feedback in regulating the baryon cycle in massive galaxies.

It can also explain the lack of cold gas in some clusters mentioned above. Those systems may represent atmospheres in which uplift is inefficient or absent, or they may be observed during a transition between feedback cycles. Although promising, detailed studies are required to assess the viability of this mechanism to explain the observed levels of molecular gas in galaxy clusters.

Uplift has become a key component of many contemporary AGN feedback models, including precipitation, stimulated cooling, and uplift-driven condensation models \citep[e.g.,][]{li14a,valentini15,mcnamara16,voit17,donahue22}. In these models, low-entropy gas lifted by AGN-inflated cavities becomes more susceptible to condensation, leading to the formation of cold gas clouds that subsequently fuel star formation and supermassive black hole (SMBH) accretion.

The dominant gas cooling mechanism in the ICM is radiation due to thermal bremsstrahlung. However, the cooling rates inferred from X-ray observations are well below those expected due to the resulting heat loss. Gas may be cooling through other channels such as by mixing with surrounding cold gas or cooling by dust grains. If the cooling X-ray gas mixes with the surrounding cold gas, it may enhance cooling and the thermal energy of the gas can be radiated at UV, optical and infrared wavelengths \citep{fabian02}. Furthermore, dust can enhance radiative cooling of the hot gas \citep{silk74}. Indeed, dust has been observed in many cool-core clusters \citep{donahue11}.

In this paper, we explore the parameter space of uplift and test whether it can account for the molecular gas observed in cool-core clusters. Using a sample of 24 clusters with well-characterized X-ray cavities and molecular gas measurements for most of them, we evaluate three key constraints on the uplift mechanism. First, we compare the mass of hot gas displaced by cavities with the observed molecular gas mass to determine whether the available displaced gas reservoir is sufficiently large for uplift to account for the cold gas. Second, we examine the velocities and heights required for gas to be uplifted and cooled within realistic dynamical timescales. Third, we estimate the amount of gas that could cool during the lifetime of the observed cavities and compare it with the molecular gas mass. Together, these tests provide quantitative constraints on the role of cavity-driven uplift in forming cold gas in cluster cores.

Throughout this paper, we have assumed a standard $\Lambda$CDM cosmology with $\Omega_{\rm m}$ = 0.3, $\Omega_\Lambda$ = 0.7, and H$_0$ = 70 km s$^{-1}$ Mpc$^{-1}$.

\section{Sample selection} \label{sec:sample}

As our goal is to study gas cooling in the inner regions of galaxy clusters, we focused primarily on systems in which multiphase gas has been identified. We selected 22 cool-core clusters with more than 90 ks of $Chandra$ observations, such that each cluster has at least one annular bin within the central 10 kpc with sufficient photon counts for spectral extraction. All clusters in our sample have central gas cooling times less than $10^9$ yr. We additionally include two cool-core clusters, MS0735+7421 and Abell 2029, which satisfy the X-ray selection criteria but lack detected atomic or molecular gas. Cold molecular gas is detected in 15 clusters, with upper limits available for five additional systems. H$\alpha$ emission is detected in 22 clusters, with one additional upper limit. The basic properties of the clusters in the sample are presented in Table~\ref{tab:clusters}.

\begin{table*}
\centering
\caption{Summary of $Chandra$ data of clusters used in our sample. Redshifts are obtained from NED and $N_H$ is obtained from \citet{hogan17b}.}
\label{tab:clusters}
\begin{tabular}{lcccclc}
\toprule
Cluster & Redshift & RA & DEC & Clean Exp & ObsIDs & $N_{\rm H}$ \\
& & & & (ks) & & (10$^{22}$ cm$^{-2}$)\\
\midrule
A1795 & 0.063001 & 13:48:52.52 & 26:35:36.30 & 760.3 & 493, 494, 3666, 5286, 5287 & 0.041 \\
& & & & & 5288, 5289, 5290, 6159, 6160 & \\
& & & & & 6161, 6162, 6163, 10898, 10899 & \\
& & & & & 10900, 10901, 12026, 12027, 12028 & \\
& & & & & 12029, 13106, 13107, 13108, 13109 & \\
& & & & & 13110, 13111, 13112, 13113, 13412 & \\
& & & & & 13413, 13414, 13415, 13416, 13417 & \\
& & & & & 14268, 14269, 14270, 14271, 14272 & \\
& & & & & 14273, 14274, 14275, 15485, 15486 & \\
& & & & & 15487, 15488, 15489, 15490, 17228 & \\
A2597 & 0.0821 & 23:25:19.72 & $-$12:07:27.62 & 625.9 & 922, 6934, 7329, 19596, 19597 & 0.0248 \\
& & & & & 19598, 20626, 20627, 20628, 20629 & \\
& & & & & 20805, 20806, 20811, 20817 & \\
A2052 & 0.0355 & 15:16:44.484 & 07:01:17.86 & 617.3 & 5807, 10477, 10478, 10479, 10480 & 0.027 \\
& & & & & 10879, 10914, 10915, 10916, 10917 & \\
Phoenix & 0.596 & 23:44:43.90 & $-$42:43:12.53 & 551.5 & 13401, 16135, 16545, 19581, 19582 & 0.015 \\
& & & & & 19583, 20630, 20631, 20634, 20635 & \\
& & & & & 20636, 20797 & \\
NGC5044 & 0.009 & 13:15:23.969 & $-$16:23:08.00 & 394.7 & 798, 9399, 17195, 17196, 17653 & 0.0487 \\
& & & & & 17654, 17666 & \\
MS0735+7421 & 0.216 & 07:41:44.21 & 74:14:38.31 & 327.4 & 4197, 10468, 10469, 10470, 10471 & 0.0328 \\
& & & & & 10822, 10918, 10922, 16275 & \\
MACS1347-11 & 0.451 & 13:47:30.58 & $-$11:45:09.21 & 293.5 & 506, 507, 2222, 3592, 13516 & 0.046 \\
& & & & & 13999, 14407 & \\
ZwCl3146 & 0.291 & 10:23:39.609 & 04:11:11.68 & 252.0 & 909, 1651, 9371 & 0.0246 \\
A1664 & 0.128 & 13:03:42.521 & $-$24:14:42.81 & 233.3 & 1648, 7901, 17172, 17173, 17557 & 0.0886 \\
& & & & & 17568 & \\
A1835 & 0.2532 & 14:01:02.08 & 02:52:42.99 & 206.3 & 495, 496, 511, 6880, 6881 & 0.02 \\
& & & & & 7370 & \\
A85 & 0.0551 & 00:41:50.48 & $-$09:18:11.82 & 195.2 & 904, 15173, 15174, 16263, 16264 & 0.0278 \\
PKS0745-191 & 0.1028 & 07:47:31.291 & $-$19:17:40.02 & 174.5 & 508, 2427, 6103, 7694, 12881 & 0.4180 \\
Hydra-A & 0.055 & 09:18:05.651 & $-$12:05:43.99 & 162.8 & 4969, 4970 & 0.043 \\
A2199 & 0.0302 & 16:28:38.245 & 39:33:04.21 & 158.3 & 497, 498, 10748, 10803, 10804 & 0.039 \\
A133 & 0.0566 & 01:02:41.59 & $-$21:52:53.65 & 141.1 & 2203, 9897, 13518 & 0.0153 \\
A262 & 0.017 & 01:52:46.482 & 36:09:06.53 & 137.4 & 2215, 7921 & 0.1250 \\
RXJ1504.1-0248 & 0.215 & 15:04:07.519 & $-$02:48:16.65 & 137.3 & 4935, 5793, 17197, 17669, 17670 & 0.0597 \\
MACS1423+24 & 0.543 & 14:23:47.87 & 24:04:42.50 & 134.1 & 1657, 4195 & 0.022 \\
A2626 & 0.0553 & 23:36:30.43 & 21:08:47.23 & 132.3 & 3192, 16136 & 0.0383 \\
IC1262 & 0.0331 & 17:33:01.973 & 43:45:35.13 & 131.7 & 2018, 6949, 7321, 7322 & 0.0178 \\
Zw2701 & 0.215 & 09:52:49.16 & 51:53:05.58 & 127.8 & 3195, 7706, 12903 & 0.00751 \\
AS1101 & 0.058 & 23:13:58.693 & $-$42:43:38.58 & 107.7 & 1668, 11758 & 0.039 \\
A2029 & 0.0773 & 15:10:56.08 & 05:44:41.05 & 107.6 & 891, 4977, 6101 & 0.033 \\
Zw7160 & 0.2578 & 14:57:15.10 & 22:20:33.89 & 92.4 & 543, 4192, 7709 & 0.0318 \\
\bottomrule
\end{tabular}
\end{table*}

\section{Data collection and Reduction}
\label{data}
\subsection{X-ray data}
The archival $Chandra$ data for clusters was obtained from the Chandra Data Archive (CDA). Data were reduced using \textsc{ciao} version 4.13 with the CalDB version 4.9.5. Level-1 events were reprocessed using the CHANDRA\_REPRO script. The VFAINT mode was turned on when observations were taken in that mode. Events with bad grades were removed and background light curves were extracted from level-2 event files on a complementary chip to the one that has most of the cluster emission. The light curves were filtered using the LC\_CLEAN routine in SHERPA to remove time intervals affected by background flares.

Blank-sky backgrounds were extracted from CalDB for each observation, reprocessed identically to the event files and normalized to a 9.5--12.0 keV count rate in each observation. Both reprocessed event files and blank-sky observations were reprojected to match the observation with the longest clean exposure time. Images were created in the 0.5--7 keV energy range for each observation ID (ObsID). Images were summed after subtracting the background. A point spread function (PSF) map of the observation with the longest exposure was created. The summed image along with the PSF map was used to detect point sources using the WAVDETECT algorithm in \textsc{ciao} \citep{freeman02}. The point source regions were visually inspected and, if necessary, adjusted in DS9 and then excluded from further analysis.

\subsubsection{Spectral extraction}
We extracted spectra in the 0.5--7 keV range from annular regions selected in such a way that each annulus has at least 3000 net counts. This criterion was chosen to ensure that the deprojected profiles have sufficient net counts for successful spectrum fitting. Although there is no hard limit to the minimum number of counts, previous studies have shown that a minimum of 3000 counts per bin leads to sufficient counts in the deprojected spectra for successful fitting. The radius of the innermost annulus was set to zero for most of the clusters. Whenever a bright point source was present in the center of the BCG, it was excluded from the innermost annulus. The center of the BCG was chosen as the center for all annuli in each cluster following \citet{hogan17a}. In most clusters at least one bin containing $\gtrsim$3000 net counts within central 10 kpc was present.

Exposure maps were created to correct the spectra for the lost area due to chip gaps and point source subtraction. The \textsc{ciao} tasks MKACISRMF and MKWARF were used to create weighted individual redistribution matrix files (RMFs) and weighted auxiliary response files (ARFs) for each spectrum. The spectra were binned to ensure 30 counts per channel. All spectra were extracted and treated separately for each ObsID. Lastly, the spectra were deprojected using the routine DSDEPROJ\footnote{\url{https://github.com/jeremysanders/dsdeproj}} described in \cite{sanders07} and \cite{russell08}.

\subsection{Molecular and H$\alpha$ gas data}
The majority of clusters in our sample are observed with ALMA. The molecular gas properties in these clusters have been discussed in detail \citep[see for example,][and references therein]{russell19,olivares19,tamhane22}. For clusters observed with ALMA, we used their molecular gas masses published in \citet{tamhane22} and \citet{russell19}. The molecular gas masses for the remaining clusters were obtained from \citet{pulido18} and \citet{edge01}. The H$\alpha$ luminosities of these clusters were obtained from \citet{hamer16} and \citet{pulido18}. The extent of the H$\alpha$ emission was taken from \citet{hamer16} and \citet{olivares19}.

\section{Results}
\label{res}

The extracted spectra and corresponding RMF and ARF files were loaded in XSPEC version 12.10.1f \citep{arnaud96} for spectral fitting. For every cluster, the spectra for all ObsIDs were loaded simultaneously for each annulus. We initially calculated projected profiles. The spectra were modeled with the absorbed single temperature (PHABS*APEC) model. The relative abundances were set to the values in \citet{anders89} and metallicities were allowed to vary. The hydrogen column density N$_{\rm H}$ was frozen to the values taken from \citet{kalberla05}. The APEC model normalization was used to calculate projected electron number densities ($n_e$) as follows:

\begin{equation}
n_e = D_{\rm A}  (1 + z) 10^7 \sqrt{\frac{N 4 \pi 1.2}{V}}
\end{equation}

where $N$ is the model normalization, $D_{\rm A}$ is the angular diameter distance to the source in cm, $z$ is the redshift of the source and $V$ is the volume of the spherical shell bounded by the inner and outer projected radius of the annulus in cm$^3$. The factor of 1.2 arises from the relative abundance of the electron to hydrogen ion number densities, $n_e$ = 1.2$n_{\rm H}$ \citep{anders89}. Assuming $n_e$ and $n_{\rm H}$ to be constant in each spherical shell, cooling times were calculated using

\begin{equation}
t_{\rm cool} = \frac{3 P}{2 n_e n_{\rm H} \Lambda(Z,T)} = \frac{3 P V}{2 L_{\rm X}}
\label{eqn:tcool}
\end{equation}

where $P$ is the pressure calculated as $P = 2 n_e k_{\rm B} T$, $\Lambda(Z,T)$ is the cooling function as a function of metallicity $Z$ and temperature $T$ and $L_{\rm X}$ is the unabsorbed bolometric X-ray luminosity obtained by integrating the fitted model between 0.1 to 100 keV by multiplying it by the CFLUX model in XSPEC. The hot gas mass in each shell was calculated using $M_{\rm gas} = PV\mu m_{\rm p}/k_{\rm B}T$, assuming that the pressure and density in each shell are constant, where $\mu$ is the mean molecular gas mass of the ICM and $m_{\rm p}$ is the mass of a proton.

\subsection{Cavities}
\label{cavities}
In this sample, X-ray cavities were detected in 21 clusters. Previous studies have examined cavities in most clusters in our sample. For those clusters, we adopted cavity shapes from the literature, and we visually inspected the remaining clusters either for more cavities or to measure the shape of cavities when only cavity distance was reported in the literature. We used the unsharp masking technique to measure the shape of the SW cavity in A1795 reported in \citet{kokotanekov18}, outer cavities in A1664 reported in \citet{calzadilla19}, and two cavities in AS1101 from their 0.5--7 keV band $Chandra$ X-ray images. To determine cavity sizes and positions, we assumed that the cavity extends to the inner edge of any surrounding bright emission. We calculated cavity volumes using the geometric mean of the oblate and prolate ellipsoidal estimates, adopting $V = \frac{4}{3}\pi (ab)^{3/2}$, where $a$ and $b$ are the projected semi-major and semi-minor axes of each cavity, respectively. To estimate the uncertainty in the intrinsic cavity volume arising from projection effects, we followed the approach described in \citet{birzan04}. Specifically, we allowed each cavity to have an intrinsic axis ratio $(a/b)$ as large as that of the most eccentric cavity observed in the sample, $(a/b)_{\max}$. The corresponding upper and lower limits on the volume were then computed by assuming either oblate or prolate symmetry for the intrinsic three-dimensional shape. In this framework, oblate geometries minimize the line-of-sight extent of the cavity, while prolate geometries maximize it, thereby bracketing the full plausible range of intrinsic volumes permitted by projection. To calculate the gas mass displaced by cavities (M$_{\rm gas, disp}$), we assumed that the cavities are empty regions devoid of any gas and multiplied the density of the ambient medium at the radius of the cavity by the cavity volume with the formula $M_{\rm gas, disp} = 1.2 n_e(R) m_{\rm p} V$. Here, $n_e(R)$ is the electron density of the ICM at the distance ($R$) of the cavity center, and $m_{\rm p}$ is the mass of a proton.
Several clusters such as A2597, A1664, and A1795 have more than two cavities indicating multiple AGN outbursts. 

To determine cavity ages, we first estimated the rise speed of each cavity at its observed projected distance from the BCG center. Two characteristic velocities were considered: the terminal velocity and the sound speed of the ambient ICM.

The terminal velocity was estimated assuming that cavities rise as buoyant bubbles through the surrounding atmosphere. Following \citet{churazov01}, the terminal velocity is

\begin{equation}
v_{\rm t} = \left( \frac{2 g V}{S\, C} \right)^{1/2},
\end{equation}

where $V$ is the cavity volume, $S = \pi ab$ is the projected cross-sectional area of the cavity, $g$ is the local gravitational acceleration interpolated from the cluster-specific profiles of \citet{hogan17b}, and $C = 0.75$ is the drag coefficient.

We also estimated the sound speed of the ambient ICM as

\begin{equation}
c_{\rm s} = \left( \frac{\gamma kT}{\mu m_{\rm p}} \right)^{1/2},
\end{equation}

where $\gamma = 5/3$ is the adiabatic index, $kT$ is the local gas temperature, $\mu = 0.61$ is the mean molecular weight, and $m_{\rm p}$ is the proton mass. The mean of the ratio between the two speeds ($v_{\rm t}/c_{\rm s}$) for the sample is 1 with the terminal speed sometimes exceeding the sound speed, due to the uncertainties in the gravitational acceleration and cavity geometries. However, physically, buoyantly rising cavities are expected to remain subsonic or transonic \citep[e.g.,][]{churazov01,birzan04}. We therefore adopt the physically motivated rise speed

\begin{equation}
v_{\rm rise} = \min(v_{\rm t},\, fc_{\rm s}),
\end{equation}

where, $f=0.5$ is based on simulations of buoyant cavities in the ICM, in which cavity terminal rise speeds are observed to be 20--50\% of the sound speeds \citep{churazov00,zhang18,zhang22}. The corresponding cavity age was then calculated as

\begin{equation}
t_{\rm age} = \frac{R}{v_{\rm rise}},
\end{equation}

where $R$ is the projected distance of the cavity from the center of the BCG. Properties of cavities are provided in table~\ref{tab:cavity_prop}.

\begin{table*}
\centering
\begin{tabular}{ccccccc}
\toprule
Cluster & a & b & R & M$_{\rm gas,\,disp}$ & $t_{\rm age}$ & Reference \\
& (kpc) & (kpc) & (kpc) & ($\times 10^8$ $\mathrm{M_{\odot}}$) & ($\times10^7$ yr) & \\
\midrule
A85 & 8.1 & 5.4 & 17.1 & 7.20$^{+12.11}_{-4.51}$ & 3.59$^{+0.80}_{-0.38}$ & \citet{shin16} \\
A133 & 9.8 & 5.2 & 28.1 & 6.55$^{+9.13}_{-3.81}$ & 6.63$^{+1.84}_{-0.15}$ & \citet{birzan04} \\
A133 & 9.8 & 5.7 & 32.7 & 6.75$^{+10.17}_{-4.06}$ & 7.92$^{+2.28}_{-0.19}$ & \citet{birzan04} \\
A262 & 2.6 & 2.6 & 6.2 & 0.65$^{+1.49}_{-0.45}$ & 2.34$^{+0.44}_{-0.03}$ & \citet{birzan04} \\
A262 & 3.3 & 2.6 & 6.7 & 0.80$^{+1.53}_{-0.52}$ & 2.45$^{+0.29}_{-0.03}$ & \citet{birzan04} \\
A1664 & 4.6 & 4.6 & 4.6 & 9.87$^{+22.57}_{-6.87}$ & 1.35$^{+0.22}_{-0.20}$ & \citet{calzadilla19} \\
A1664 & 4.6 & 4.6 & 6.7 & 10.94$^{+25.01}_{-7.61}$ & 2.25$^{+0.43}_{-0.29}$ & \citet{calzadilla19} \\
A1664 & 10.1 & 7.5 & 15.0 & 31.13$^{+56.97}_{-20.13}$ & 4.36$^{+0.59}_{-0.79}$ & This work$^*$ \\
A1664 & 9.8 & 7.2 & 23.3 & 18.11$^{+32.81}_{-11.67}$ & 6.05$^{+0.78}_{-0.77}$ & This work$^*$ \\
A1795 & 10.0 & 10.0 & 44.0 & 26.53$^{+60.65}_{-18.46}$ & 9.01$^{+4.09}_{-0.20}$ & \citet{kokotanekov18} \\
A1795 & 23.0 & 7.0 & 16.0 & 91.50$^{+74.36}_{-41.02}$ & 3.51$^{+0.36}_{-0.38}$ & \citet{kokotanekov18} \\
A1795 & 12.3 & 8.3 & 37.6 & 29.43$^{+49.85}_{-18.51}$ & 7.37$^{+2.46}_{-0.21}$ & This work$^*$ \\
A1835 & 15.5 & 11.6 & 23.3 & 236.80$^{+436.29}_{-153.49}$ & 4.51$^{+0.14}_{-0.13}$ & \citet{mcnamara06} \\
A1835 & 13.6 & 9.7 & 16.6 & 175.12$^{+310.82}_{-112.01}$ & 3.60$^{+0.14}_{-0.14}$ & \citet{mcnamara06} \\
A2052 & 6.5 & 6.0 & 6.7 & 9.73$^{+20.99}_{-6.65}$ & 2.00$^{+0.15}_{-0.01}$ & \citet{birzan04} \\
A2052 & 10.7 & 7.8 & 11.2 & 24.72$^{+44.62}_{-15.91}$ & 3.34$^{+0.01}_{-0.01}$ & \citet{birzan04} \\
A2199 & 6.5 & 6.5 & 18.9 & 5.53$^{+12.64}_{-3.85}$ & 4.50$^{+1.85}_{-0.22}$ & \citet{birzan04} \\
A2199 & 6.2 & 3.5 & 21.2 & 1.87$^{+2.75}_{-1.12}$ & 4.85$^{+2.65}_{-0.21}$ & \citet{birzan04} \\
A2597 & 7.3 & 6.0 & 21.7 & 13.17$^{+26.06}_{-8.75}$ & 4.96$^{+2.11}_{-0.19}$ & \citet{shin16} \\
A2597 & 3.1 & 2.9 & 9.4 & 2.20$^{+4.80}_{-1.51}$ & 2.75$^{+1.04}_{-0.11}$ & \citet{shin16} \\
A2597 & 8.7 & 6.6 & 21.6 & 19.85$^{+36.95}_{-12.91}$ & 4.93$^{+1.50}_{-0.19}$ & \citet{shin16} \\
A2597 & 5.1 & 3.0 & 20.6 & 2.85$^{+4.33}_{-1.72}$ & 5.00$^{+2.94}_{-0.47}$ & \citet{shin16} \\
A2597 & 14.0 & 9.6 & 17.4 & 86.48$^{+148.81}_{-54.69}$ & 4.03$^{+0.16}_{-0.17}$ & \citet{shin16} \\
A2626 & 19.9 & 12.4 & 18.1 & 68.34$^{+108.91}_{-41.99}$ & 4.23$^{+0.29}_{-0.29}$ & \citet{shin16} \\
AS1101 & 4.6 & 3.5 & 5.8 & 3.40$^{+6.27}_{-2.20}$ & 1.80$^{+0.06}_{-0.07}$ & This work$^*$ \\
AS1101 & 4.6 & 3.5 & 11.2 & 2.95$^{+5.44}_{-1.91}$ & 3.01$^{+0.66}_{-0.10}$ & This work$^*$ \\
Hydra-A & 20.5 & 12.4 & 24.9 & 114.65$^{+178.33}_{-69.78}$ & 5.46$^{+0.16}_{-0.16}$ & \citet{wise07} \\
Hydra-A & 21 & 12.3 & 25.6 & 114.49$^{+173.41}_{-68.96}$ & 5.60$^{+0.16}_{-0.16}$ & \citet{wise07} \\
Hydra-A & 47.2 & 31.5 & 100.8 & 381.01$^{+641.70}_{-239.07}$ & 20.46$^{+0.26}_{-0.26}$ & \citet{wise07} \\
Hydra-A & 29 & 20.9 & 59.3 & 216.73$^{+387.81}_{-139.03}$ & 11.80$^{+0.14}_{-0.14}$ & \citet{wise07} \\
Hydra-A & 105 & 99.7 & 225.6 & 2858.84$^{+6294.34}_{-1965.93}$ & 43.40$^{+0.31}_{-0.34}$ & \citet{wise07} \\
Hydra-A & 67.7 & 50.1 & 104.3 & 1264.92$^{+2310.42}_{-817.40}$ & 21.29$^{+0.26}_{-0.26}$ & \citet{wise07} \\
IC1262 & 2.2 & 1.5 & 6.5 & 0.98$^{+2.23}_{-0.68}$ & 2.68$^{+0.93}_{-0.56}$ & \citet{pandge19} \\
IC1262 & 4.0 & 2.0 & 6.1 & 1.07$^{+2.45}_{-0.75}$ & 3.94$^{+1.08}_{-0.73}$ & \citet{pandge19} \\
MACS1423+24 & 27.1 & 12.7 & 36.1 & 387.71$^{+484.36}_{-215.34}$ & 6.15$^{+0.22}_{-0.23}$ & \citet{shin16} \\
MACS1423+24 & 9.4 & 9.4 & 15.8 & 105.40$^{+240.92}_{-73.32}$ & 3.11$^{+0.11}_{-0.11}$ & \citet{diehl08} \\
MACS1423+24 & 9.4 & 9.4 & 16.7 & 101.89$^{+232.90}_{-70.88}$ & 3.27$^{+0.12}_{-0.12}$ & \citet{diehl08} \\
MS0735+7421 & 109.0 & 106.0 & 150.0 & 7243.03$^{+16225.71}_{-5007.65}$ & 26.05$^{+1.39}_{-1.21}$ & \citet{vantyghem14} \\
MS0735+7421 & 120.0 & 100.0 & 186.0 & 6189.15$^{+12374.79}_{-4125.71}$ & 29.39$^{+1.11}_{-1.13}$ & \citet{vantyghem14} \\
NGC5044 & 3.1 & 1.7 & 6.1 & 0.34$^{+0.49}_{-0.20}$ & 2.72$^{+0.03}_{-0.03}$ & \citet{gastaldello09} \\
NGC5044 & 3.1 & 1.7 & 6.1 & 0.34$^{+0.49}_{-0.20}$ & 2.72$^{+0.03}_{-0.03}$ & \citet{gastaldello09} \\
Phoenix & 11.3 & 8.6 & 17.3 & 248.72$^{+464.22}_{-161.95}$ & 2.31$^{+0.12}_{-0.13}$ & \citet{mcdonald19} \\
Phoenix & 14.6 & 8.6 & 17.3 & 365.28$^{+555.87}_{-220.43}$ & 2.31$^{+0.12}_{-0.13}$ & \citet{mcdonald19} \\
PKS0745 & 9.9 & 8.8 & 18.4 & 68.67$^{+144.05}_{-46.50}$ & 3.89$^{+0.11}_{-0.10}$ & \citet{shin16} \\
PKS0745 & 11.6 & 8.5 & 9.3 & 145.54$^{+263.81}_{-93.80}$ & 2.33$^{+0.11}_{-0.10}$ & \citet{shin16} \\
RXCJ1504 & 10.0 & 7.0 & 10.0 & 131.32$^{+229.68}_{-83.55}$ & 1.95$^{+0.16}_{-0.16}$ & \citet{vantyghem18} \\
Zw2701 & 43.5 & 43.5 & 54.1 & 1693.89$^{+3871.76}_{-1178.36}$ & 9.83$^{+0.37}_{-0.36}$ & \citet{diehl08} \\
Zw2701 & 36.7 & 36.7 & 48.6 & 1114.30$^{+2546.97}_{-775.17}$ & 8.61$^{+0.33}_{-0.33}$ & \citet{diehl08} \\
ZwCl3146 & 33.2 & 33.2 & 39.8 & 1760.00$^{+4022.86}_{-1224.35}$ & 6.98$^{+0.53}_{-0.54}$ & \citet{diehl08} \\
ZwCl3146 & 33.0 & 33.0 & 59.1 & 1305.11$^{+2983.12}_{-907.91}$ & 10.55$^{+0.50}_{-0.53}$ & \citet{diehl08} \\
\bottomrule
\end{tabular}
\caption{Properties of cavities detected in clusters in our sample. \\$^*$: Only distances for these cavities were reported in the literature. Therefore, we measured their shapes and other properties as described in section~\ref{cavities}.}
\label{tab:cavity_prop}
\end{table*}

\section{Discussion}
Understanding the uplift of low entropy gas by buoyantly rising X-ray cavities is crucial in studying the intricate dynamics of the radio-mode AGN feedback. This mechanism is expected to play an important role in rapid gas cooling in central regions of galaxy clusters. However, its effectiveness has not been explored in detail. In this section, we discuss the uplift mechanism in the context of the $t_{\rm cool}/t_{\rm ff}$ and $t_{\rm cool}/t_{\rm I}$ thresholds. Assuming that these thresholds are the determining physics for gas cooling, we explore the parameter space of gas uplift velocity, lifting radius and IR cooling rate and compare it with observations.

\subsection{How much gas is in central regions?}

\begin{figure}
    \centering
    \includegraphics[width=0.5\textwidth]{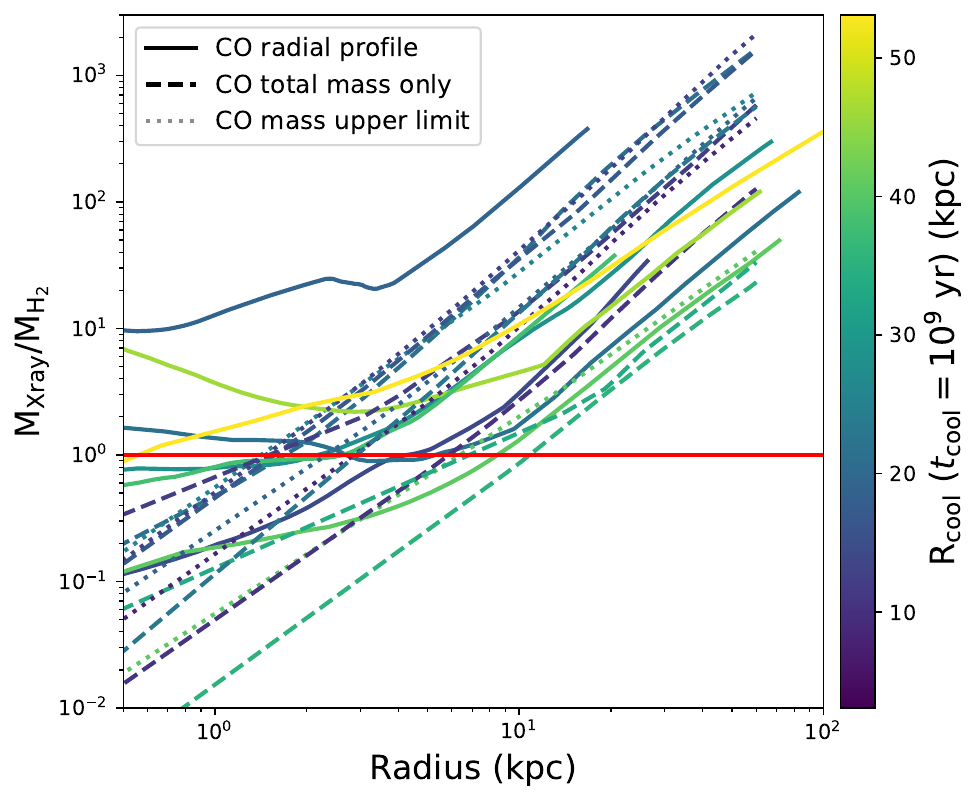}
    \caption{Radial dependence of the ratio of hot to cold gas mass in the central regions of clusters in our sample. The red horizontal line shows a ratio of unity. The clusters are colored by the cooling radius defined as the radius at which the cooling time is 10$^9$ yrs.}
    \label{fig:1}
\end{figure}

The stimulated feedback scenario proposes that low-entropy gas near the cluster center is uplifted by AGN-inflated cavities to radii where its cooling time becomes shorter than its infall time, allowing it to condense before returning to the core. For cavities to lift this gas, a sufficiently large reservoir of low-entropy material must be available. To assess this, we computed the ratio of hot, X-ray emitting gas mass to the total cold molecular gas mass within a given radius. Hot gas masses were obtained by interpolating (or, beyond the radial coverage of the projected profile, extrapolating) each cluster's projected gas mass profile. For the subset of clusters in common with the \citet{tamhane22} sample for which a CO moment-0 map was available, we generated annular regions centered on the continuum peak and extracted the flux profile in these annuli. This profile was then scaled by the total molecular gas mass of the cluster to obtain a molecular gas mass profile. We interpolated the hot gas mass onto the same radial grid as this profile and computed the ratio point by point. For the remaining clusters, where only a total molecular gas mass or only upper limits are available, we assumed this mass to be constant with radius between 1 and 30 kpc and divided the hot gas mass profile by this single value. The constant-mass assumption is motivated by the fact that most molecular gas in cool-core systems lies within the central $\sim$10 kpc \citep[e.g.,][]{russell19,tamhane22}. It introduces at most a factor of 2 uncertainty in the ratio for these clusters in the inner regions and does not affect our conclusions qualitatively.

Figure~\ref{fig:1} shows the ratio of the hot to cold gas mass as a function of radius. In nearly all clusters, the mass of the hot gas exceeds the mass of the cold gas within the central $\sim10$ kpc, indicating that cluster cores contain a sufficiently large hot gas reservoir to account for the observed cold gas without requiring substantial external accretion. Beyond 10~kpc, hot gas dominates the mass budget in every cluster. The curves are color-coded by each cluster's cooling radius, $R_{\rm cool}$ (the radius where $t_{\rm cool} = 1$ Gyr), and there is a rough trend where $M_{\rm Xray}/M_{\rm H_2} = 1$ to shift outward in clusters with a larger cooling radius, consistent with the ratio profile scaling with the physical extent of the cool core, although with considerable scatter about this trend.

However, this comparison does not imply that all of the central hot gas will cool into the cold phase. Only a fraction of the hot gas is expected to be lifted by cavities, and only part of that lifted gas may cool efficiently before returning. Consequently, the observed cold gas reservoir may represent only a subset of the gas influenced by uplift over multiple feedback cycles. In the following sections, we estimate whether cavity-driven lifting can account for the observed cold gas masses.

\subsection{How much gas can cavities lift?}
\label{sec:lifting}

\begin{figure}
    \centering
    \includegraphics[width=0.53\textwidth]{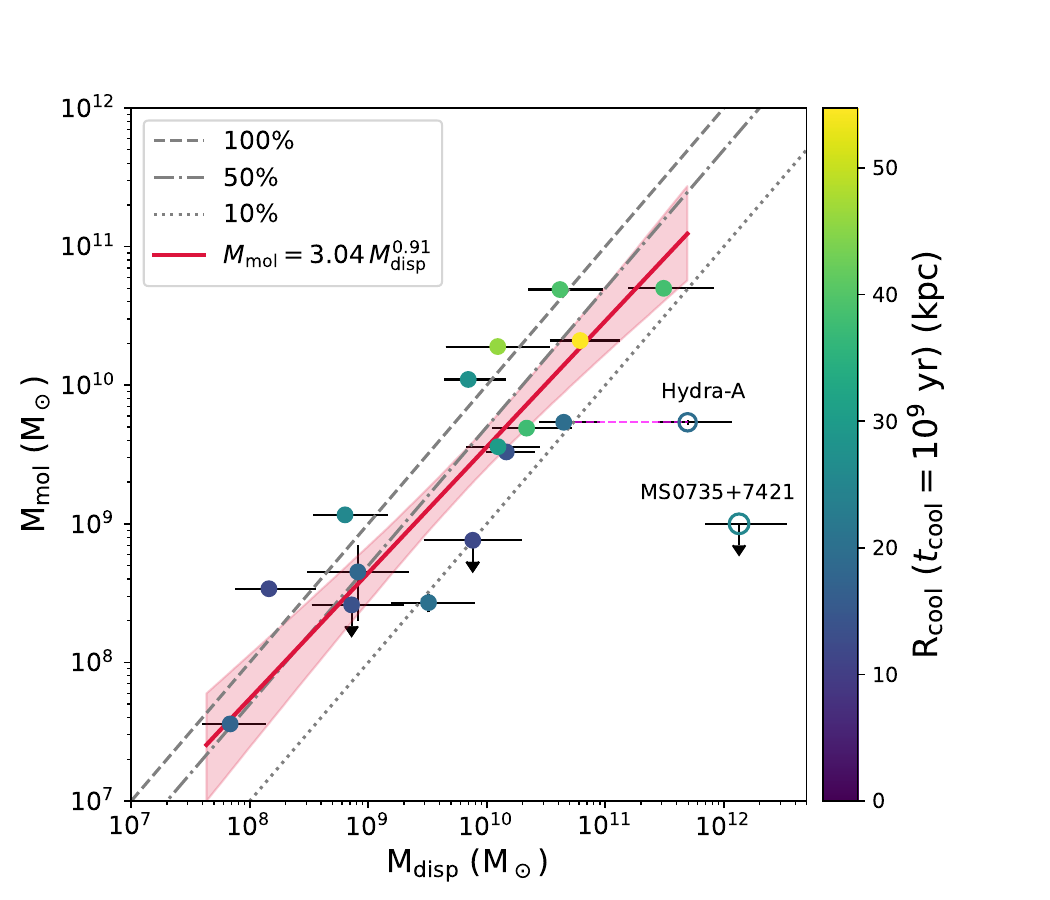}
    \caption{The relationship between the total molecular gas mass in BCGs and the total maximum displaced gas mass by cavities. In many clusters, the mass of hot gas displaced by cavities is comparable to the observed mass of cold gas. The dashed, dash-dotted and dotted lines indicate levels where the observed molecular gas mass is 100, 50 and 10 per cent of the maximum displaced gas mass, respectively. The red line and red shaded region shows the best-fit relation excluding outer cavities and the corresponding 1$\sigma$ error interval. The hollow circles show displaced gas masses containing outer very large cavities are included. The magenta dashed line connects the points representing Hydra-A including and excluding outer cavities. The points are colored by the cooling radius.}
    \label{fig:mdisp}
\end{figure}

We compare the mass of hot gas displaced by cavities with the total molecular gas mass in galaxy clusters. Molecular gas represents the endpoint of cooling processes and typically contains most of the mass in both inflows and outflows detected in galaxies \citep[e.g.,][]{olivares19,russell19}. It also directly fuels star formation and central supermassive black holes and therefore has the greatest impact on galaxy evolution. In a previous study, \citet{tamhane22} showed that the kinetic power of molecular outflows in BCGs is typically only a few percent of the cavity mechanical power, demonstrating that radio bubbles are energetically capable of driving the observed cold gas outflows. Here, we address the complementary question of whether the observed cavities can physically displace enough low-entropy gas to account for the molecular gas reservoirs. The cavity sizes used to estimate the displaced gas masses are listed in Table~\ref{tab:cavity_prop}, and the method used to estimate the displaced gas mass is described in Section~\ref{cavities}.

Several studies of individual galaxy clusters have shown that the mass of hot gas lifted by cavities can exceed the cold gas mass observed in galaxy clusters \citep[see, for example,][]{mcnamara14,russell17b,vantyghem18}. We systematically investigate whether the observed cold gas masses are comparable to the displaced gas masses by cavities produced during single or multiple bursts of AGN activity. The amount of gas that cavities can actually lift is limited by the weight of the gas they displace, as described by Archimedes' principle. Thus, cavities cannot lift more gas than the mass of gas they displace. The displaced gas mass therefore provides a natural upper bound on the amount of gas that can be directly lifted or entrained by the cavities.

In Fig.~\ref{fig:mdisp}, we compare the total hot gas mass displaced by all detected cavities in a cluster $M_{\rm disp}$ with the total molecular gas mass $M_{\rm mol}$ detected in that cluster. When multiple generations of cavities are present in a cluster (e.g., in A1664, A1795, A2597, Hydra-A and MACS1423+24), we sum their displaced gas masses to estimate the cumulative displaced gas mass. In twelve clusters, $M_{\rm disp}$ exceeds $M_{\rm mol}$. For Hydra-A, the mass displaced from only inner cavities is also shown, which was more closely correlated with the molecular gas mass as discussed in the following section and further in Section~\ref{sec:exceptions}. In A1795, A2597 and Hydra-A the total gas mass displaced by all cavities exceeds the total observed molecular gas mass, indicating that the cumulative gas mass displaced by repeated AGN outbursts is sufficiently large, in principle, to account for the present day molecular gas reservoir.

In A262, A1664, A1835, AS1101, and RXCJ1504, the molecular gas mass exceeds the maximum gas mass that can be displaced by the observed cavities. This may indicate that the molecular gas in these clusters accumulated and survived over multiple AGN feedback cycles, was acquired externally, or formed through additional processes. Additionally, sloshing features and evidence for a possible merger have also been reported in A1664 \citep{calzadilla19}, and such dynamical disturbances could contribute additional cold gas or affect the lifting efficiencies. In RXCJ1504, only a single cavity has been detected \citep{vantyghem18}. If an undetected complementary cavity is present, the true displaced gas mass would be larger, reducing or potentially eliminating the discrepancy in this cluster.

Across the full sample, the median and mean ratios of $M_{\rm mol}/M_{\rm disp}$ are 0.35 and 0.72, respectively, excluding the large outer cavities in MS0735+7421 and Hydra-A. The $1\sigma$ range extends from approximately 0.14 to 1.56. If roughly half of the molecular gas resides in extended filaments \citep{russell19,tamhane22}, these ratios decrease by a factor of two, giving a median of $\sim0.18$ and a mean of $\sim0.36$. For the 12 systems with $M_{\rm mol}<M_{\rm disp}$, the mean ratio is 0.27 for the total molecular gas and 0.14 for the filamentary component. Thus, if all uplifted gas eventually cools to the molecular phase, these systems would require on average about 27\% of the displaced gas to produce the total molecular reservoir, or about 14\% to produce only the filamentary component. \citet{pope10} estimate that the mass of upwardly displaced gas associated with a spherical bubble is approximately half of the mass displaced by the bubble. Under this assumption, the observed total and filamentary molecular reservoirs would correspond to about 54\% and 28\%, respectively, of the upwardly displaced gas.

We also fit a linear relationship between log $M_{\rm mol}$ and log $M_{\rm disp}$ using {\sc linmix}\footnote{\url{https://github.com/jmeyers314/linmix}} \citep{kelly07} which accounts for errors in both variables and upper limits on $M_{\rm mol}$. Excluding MS0735+7421 and outer cavities in Hydra-A, which is an extreme outlier, we obtain,

\begin{equation}
    {\rm log} \, M_{\rm mol} = 0.91^{+0.17}_{-0.16} \,\, {\rm log} \, M_{\rm disp} + 0.49^{+1.52}_{-1.61}
\end{equation}
and an intrinsic scatter of 0.47 dex. The relation indicates that the molecular gas mass scales approximately linearly with the displaced gas mass across the sample, when only cavities with radii $<60$ kpc are included. However, the substantial intrinsic scatter suggests that additional factors influence the final molecular gas reservoir which may include changes in lifting and cooling efficiency, cavity duty cycle, dynamical state of the cluster and contribution from other cooling processes.

Overall, these results indicate that cavities satisfy both the energetic requirements for driving molecular outflows and the mass budget requirements for lifting low entropy gas capable of cooling into molecular clouds. Most systems are consistent with the molecular gas having formed from gas lifted by the observed cavities, while the remaining systems likely require contributions from multiple AGN feedback cycles, external gas acquisition, or other processes.

Both molecular gas and displaced gas masses have large uncertainties. The dominant uncertainty in the displaced gas mass arises from the estimation of the cavity volume. In some clusters, determining the exact boundary of a cavity is difficult, and some cavities may have partially collapsed. In addition, the projected cavity shape may either under- or overestimate the true cavity volume. Molecular gas masses may also be over- or underestimated if the CO-to-H$_2$ conversion factor ($X_{\rm CO}$) differs from the Galactic value. The typical $X_{\rm CO}$ has an uncertainty of a factor of a few \citep{bolatto13}. Improved multiwavelength observations of cavities and of the physical conditions in molecular gas will be required to reduce these uncertainties.

\subsubsection{Exceptional cases of MS0735+7421 and Hydra-A}
\label{sec:exceptions}

MS0735+7421 and Hydra-A host some of the largest cavities in our sample extending to large cluster centric radii ($R>60$ kpc). In MS0735+7421, the available molecular gas upper limit suggests that cold gas in this system is very small compared to the size of its cavities. Hydra-A similarly has large cavities, implying a very high maximum displaced gas mass, placing the system well below the mean $M_{\rm mol}$--$M_{\rm disp}$ relation when large cavities are included and much closer to the mean relation when they are excluded from total displaced gas mass. Two effects may contribute to this. First, is that the present day properties of these large, distant and old cavities do not accurately reflect their role at earlier stages of their evolution. As cavities rise through the atmosphere, they expand as the ambient pressure decreases, so the displaced gas mass inferred from their current volume can be substantially larger than when the cavity was younger and closer to the cluster center. Gas uplifted during these earlier stages may also dynamically decouple from the rising cavity, condense and fall back as the cavity continues to rise. In this case, the present day $M_{\rm disp}$ of an old cavity may not provide a reliable measure of the amount of low-entropy gas that it was capable of lifting during the phase most relevant for condensation.

Alternatively, the most energetic or spatially extended cavity systems may not necessarily contribute substantially to the formation of cold gas, perhaps because they couple inefficiently to the low-entropy gas in the central regions where condensation is most likely to occur. One possibility is that these cavities may be formed by short-lived but powerful jets or jets with a large jet base such that they overheat the central regions of the cluster, delaying the formation of cold gas filaments, as indicated by some simulations \citep[e.g.,][]{duan24}. In contrast, highly collimated jets with narrow opening angles may drill through the central atmosphere and deposit most of their energy at larger radii, which may not couple to gas within the inner $\sim10$ kpc region \citep[e.g.,][]{prasad22}. This process can inflate cavities at larger distances, which may not be efficient at lifting low-entropy gas from the center. More generally, displaced gas mass is an upper limit for the total amount of gas available for uplift. Even if substantial uplift of the gas occurs, condensation may remain inefficient if the lifted material has relatively high entropy or is lifted at very high speeds \citep[e.g.,][]{voit17}. These systems therefore emphasize that uplift-driven condensation may depend not only on cavity size but also on cavity evolution, cavity--ICM coupling, the thermodynamic state of the uplifted gas, and the radius at which the jet deposits its energy.

\subsubsection{Contribution of star formation to cooling mass budget}
While the molecular gas reservoir represents the current endpoint of cooling, a complete mass budget should also include the stellar mass formed from this gas. For clusters with available star formation rates (Table~\ref{tab:cluster_properties}), we estimate the stellar mass formed assuming a continuous star formation history over a characteristic timescale of $\sim 10^8$ yr. Under this assumption, the inferred stellar masses correspond to $\sim 15 \pm 12$\% of the observed total molecular gas mass, excluding the Phoenix cluster, whose exceptionally high star formation rate yields a stellar mass comparable to several times its present molecular gas reservoir. 

The inferred fractions span a broad range ($\sim$1--45\%), reflecting significant system-to-system variation. In moderate to highly star forming systems such as A1795, A1835 and RXCJ1504, the percentage ranges from 30--45\%. Excluding these systems as well, results in a stellar mass corresponding to only 10\% of the total molecular gas mass in BCGs. In most clusters, the stellar mass formed is therefore subdominant compared to the molecular gas reservoir. However, in a few systems with elevated star formation rates, the contribution can be significant.

These estimates are likely upper limits, since star formation in BCGs is generally episodic rather than continuous over $10^8$ yr timescales. Consequently, including the stellar mass formed from cooled gas increases the total cooled mass only modestly for the majority of systems and does not qualitatively alter our comparison between the molecular gas reservoir and the maximum gas mass that cavities can displace.

\subsection{What speeds and heights are necessary for lifting gas?}
\label{sec:lifting_vel_rad}

\subsubsection{Kinematics of Cavity-Driven Lifting}

Due to the lack of direct high-resolution observations, the lifting speeds of low-entropy gas are not definitively constrained. However, it is expected that the gas initially couples with the cavity-driven flow, moving at speeds comparable to the buoyant rise speeds of the cavities, typically a few hundred km s$^{-1}$ \citep[see, e.g.,][]{gingras24,reefe25a}. In one of the first observations, XRISM has revealed velocities of $\sim 150$ km s$^{-1}$ in the wake of a potential giant cavity in the Ophiuchus cluster \citep{russell26}. As the gas rises, it gradually decouples due to drag and gravity, eventually slowing to velocities comparable to the bulk motions of the hot ICM \citep[e.g.,][]{hitomi18,xrism_a2029,xrism_centaurus}.

To understand the conditions conducive to gas condensation, classical models often rely on the ratio of the cooling time to the free-fall time ($t_{\rm c}/t_{\rm ff} \lesssim 10-30$) \citep[e.g.,][]{mccourt12, voit17}. However, cooling gas also experiences drag forces and ram pressure. Therefore, following \citet{mcnamara16}, we use the infall time ($t_{\rm I}$), which accounts for the terminal velocity of decoupled clouds and serves as a more accurate timescale than the free-fall time.

We explore this parameter space by assuming the gas is initially lifted with a mean speed $v_{\rm lifting}$ between 100 and 400 km s$^{-1}$, reflecting typical observed speeds of cold gas and cavity rise speeds. After decoupling, the gas falls back at its terminal velocity, $v_{\rm term}$. The total time for the gas to be lifted to a radius $R$ and return to its initial height, which we denote as $t_{\rm I}$ is:
\begin{equation}t_{\rm I} = \frac{R}{v_{\rm lifting}} + \frac{R}{v_{\rm term}},
\label{eq:tI}
\end{equation}
where we estimate the lifting radius $R$ by setting $t_{\rm I} = t_{\rm cool}$, where $t_{\rm cool}$ is determined using Equation~\ref{eqn:tcool}.

\subsubsection{Comparing Lifting Models to Observations}

\begin{figure*}
    \includegraphics[width=1.09\textwidth]{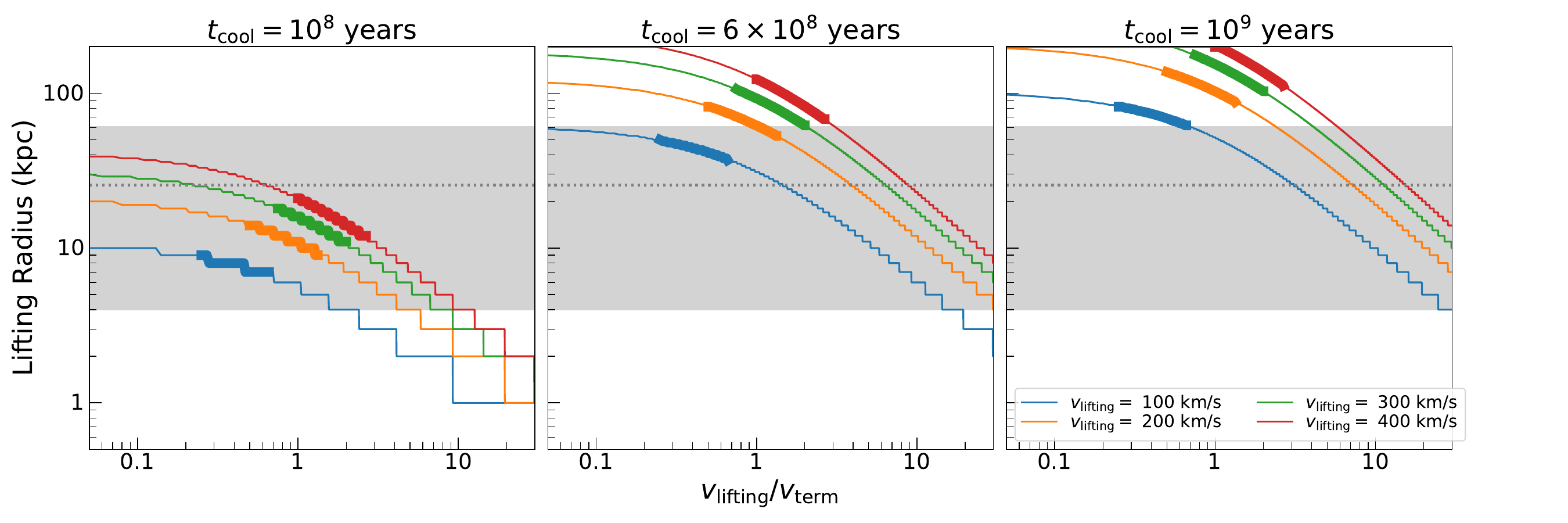}
    \caption{The figure illustrates the relationship between uplift radius ($R$) and the ratio of uplift velocity to terminal velocity for gas cooling times of 10$^8$, 6$\times10^8$ and 10$^9$ years, shown in the left, middle and right panels, respectively. The colored lines indicate different uplift velocities, ranging from 100 to 400 km s$^{-1}$, as indicated in the legend in the right panel. The bold regions of each line indicates the range of ratios where terminal velocities lie in the range between 150 and 400 km s$^{-1}$. The shaded grey region shows the observed range of maximum extent of the H$\alpha$ emission. The dashed grey line indicates the average of the largest extent of H$\alpha$ emission in BCGs in this sample of 30 kpc.}
    \label{fig:rad-vel}
\end{figure*}

Figure~\ref{fig:rad-vel} displays the relationship between the lifting radius and the velocity ratio ($v_{\rm lifting}/v_{\rm term}$) for various cooling times. The bold regions highlight the physical parameter space where $v_{\rm lifting}$ is 100--400 km s$^{-1}$ and $v_{\rm term}$ is 150--400 km s$^{-1}$. We adopt 150 km s$^{-1}$ as a representative minimum terminal velocity scale, motivated by the characteristic line-of-sight velocity dispersions of $\sim$100--170 km s$^{-1}$ measured in cool-core atmospheres with XRISM \citep[e.g.,][]{xrism_a2029,mcnamara26}.

We use the spatial extent and kinematics of the warm ionized gas to compare with the inferred lifting radii and velocities. The warm gas is particularly useful for this purpose because it is typically more spatially extended than the molecular gas and traces the spatial distribution and kinematics of the multiphase filaments associated with the cooling ICM. In contrast, the molecular gas provides a more appropriate comparison for the total mass of gas that has cooled. For gas with $t_{\rm cool} \sim 10^8$ yr, lifting velocities of 150--400 km s$^{-1}$, comparable to those observed in the warm and molecular phases, correspond to lifting radii of $\sim$4--20 kpc, well within the observed extent of the warm filaments ($\lesssim 60$ kpc). For the sample average cooling time of $t_{\rm cool} \sim 6 \times 10^8$ yr, the corresponding lifting radii increase to $\sim$40--100 kpc, approaching or exceeding the observed extent of the warm gas in many systems. Gas with $t_{\rm cool} \sim 10^9$ yr would require even larger lifting radii of $\sim$60--200 kpc.

In reality, the lifted gas is likely multiphase, spanning a range of temperatures, with pockets of denser gas with shorter cooling
times intermixed with slightly more diffuse, slower cooling gas. As a result, some gas may begin cooling early during the lifting phase, while other parcels may cool at late times or on their return toward the cluster center. This leads to the presence of warm gas at a wide range of radii, consistent with the observed range of warm gas radii in BCGs. If gas with longer cooling times is responsible for the warm gas at 10--30 kpc, it must have been lifted to large radii and begun cooling during its infall phase. Extended metal-enriched outflows extending several hundred kiloparsecs from the centers of BCGs in clusters with cavities inflated by radio jets support the plausibility of such large lifting radii \citep{nulsen05,simionescu08,simionescu09,werner10,kirkpatrick15}. Overall, explaining the observed warm/cold gas distributions and kinematics is most straightforward if the initial low-entropy gas lifted by cavities had relatively short cooling times ($\lesssim$ a few $\times 10^8$ years).

\subsubsection{Turbulent Gas Transport}

\begin{figure}
    \includegraphics[width=0.48\textwidth]{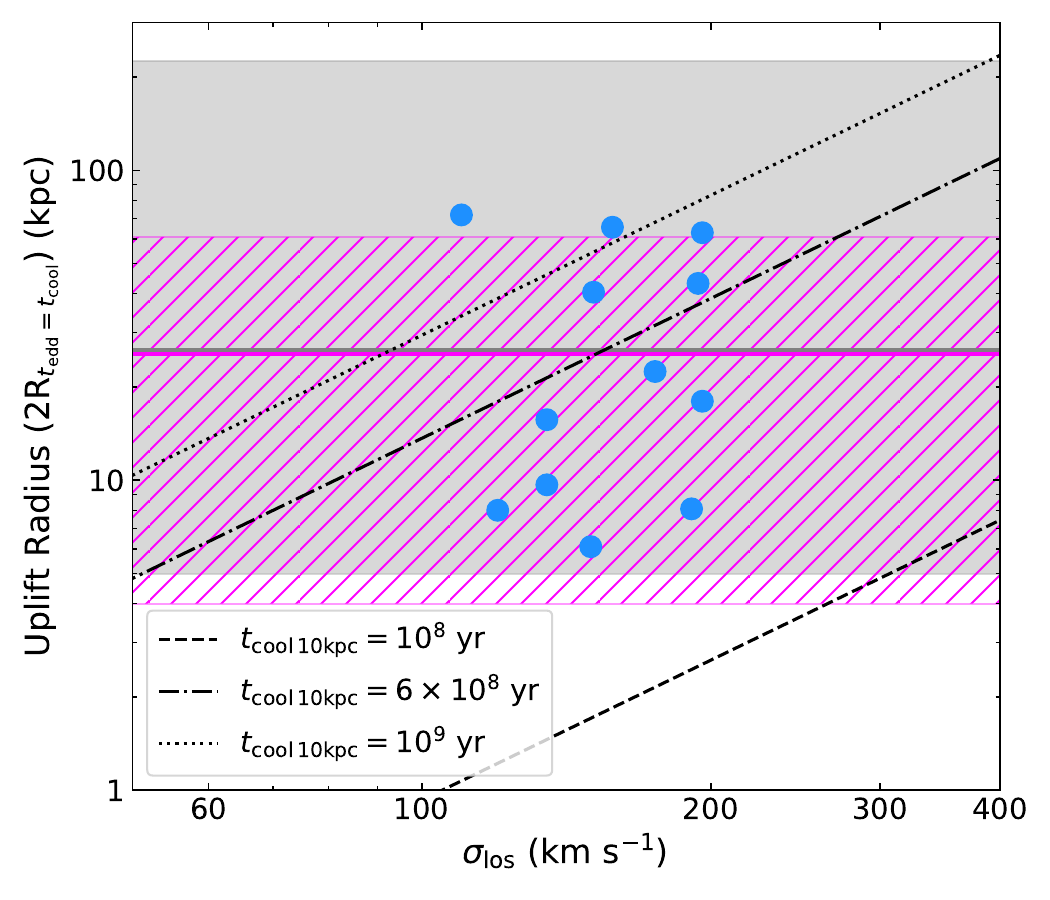}
    \caption{Figure shows the characteristic turbulent transport scale, taken here as the diameter of circular turbulent eddies, as a function of velocity dispersion for cooling gas parcels. The turbulent transport scale is calculated by requiring $t_{\rm eddy}=t_{\rm cool}$, where $t_{\rm eddy}$ is the eddy turnover timescale and $R$ in the turnover-time relation corresponds to the eddy radius. The black lines show the corresponding eddy diameters calculated for the sample mean $t_{\rm cool,\,10\,kpc}\sim6\times10^8$ yr and for $t_{\rm cool,\,10\,kpc}=10^8$ and $10^9$ yr for a cavity of diameter 20 kpc. The grey shaded region and solid grey line indicate the range and mean of the cavity radii in the sample, respectively. The magenta shaded region and solid magenta line similarly indicate the range and mean of the maximum extent of the H$\alpha$ emission.}
    \label{fig:vt_RL}
\end{figure}

The previous subsection considered approximately linear transport of low-entropy gas. However, gas may also be transported by turbulent motions in the hot atmosphere. Figure~\ref{fig:vt_RL} therefore shows the characteristic spatial scale associated with turbulent eddies, following the framework of \citet{gaspari18} and \citet{olivares19}. For an eddy of characteristic radius $R$, the turnover time is
\begin{equation}
t_{\rm eddy} = 2 \pi \frac{R^{2/3}L^{1/3}}{\sigma_v},
\end{equation}
where $\sigma_v$ is the velocity dispersion at the injection scale $L$. For individual systems, we adopt the major axis of the largest cavity as the characteristic injection scale. We excluded large cavities in Hydra-A and MS0735+7421 as discussed in Section~\ref{sec:lifting}. 

For a given cooling time, we solve $t_{\rm eddy}=t_{\rm cool}$ for $R$. The resulting $R$ gives a characteristic radius of the turbulent eddy rather than the maximum radial height reached by gas uplifted from the galaxy center. However, its comparison with the observed H$\alpha$ extent provides an approximate test of whether turbulent motions can redistribute low-entropy gas over spatial scales comparable to those of the multiphase filaments before the gas cools.

Theoretical tracks are shown for cooling times of $10^8$, $6\times10^8$, and $10^9$ yr for a typical cavity, with a diameter of 20 kpc, which is close to the average of major axes of cavities considered here. Gas with relatively short initial cooling times, $\lesssim$ a few $\times10^8$ yr, is associated with turbulent transport scales comparable to the characteristic extents of the observed H$\alpha$ filaments and cavities, whereas gas with longer cooling times requires correspondingly larger eddy scales.

Overall, both the linear uplift and turbulent transport models show the range of velocities, cooling times, and spatial scales over which low-entropy gas can be transported before cooling. In both cases, gas with relatively short initial cooling times, $\lesssim$ a few $\times10^8$ yr, can reach distances comparable to the observed H$\alpha$ filament extents for characteristic velocities of a few hundred km s$^{-1}$. Gas with longer cooling times generally needs to be lifted to much larger distances, which are less consistent with the typical cavity sizes and H$\alpha$ extents in the sample.

\subsection{Inflow of high entropy gas}
\label{preagntc}

When cavities are inflated by radio jets and rise through the ICM, gas from larger radii, which typically has higher entropy and longer cooling times, is redistributed and partially flows inward to refill the displaced volume and maintain an approximate pressure equilibrium. This process occurs on timescales of $10^7$--$10^8$ years, comparable to or up to about twice the age of the observed cavities \citep[see, for example,][]{birzan04,rafferty06}. As a result, the gas currently observed in the inner regions is not necessarily representative of the gas that was originally present prior to the AGN outburst. In particular, the inflow of higher-entropy gas is expected to increase the cooling time and entropy while reducing the gas density in the central $\sim$5--10 kpc region by factors of $\sim$2--4 \citep{soker02,prasad22}. 

Prior to the outburst, the central atmosphere may therefore have been denser and had a shorter cooling time than inferred from the present day profiles. To estimate the maximum possible effect of this mechanism, we consider an extreme limiting case in which all of the gas displaced by the cavities originated within the inner 10 kpc, was uplifted from this region, and has not subsequently been replenished by inflow from larger radii. We therefore add the displaced gas mass to the currently observed gas mass within 10 kpc to obtain an upper limit on the pre-outburst density. Assuming approximate pressure equilibrium, for which $t_{\rm cool}\propto n^{-3/2}$, the inferred pre-outburst cooling times are shorter by a factor of $\sim2$ on average for the sample after excluding systems with exceptionally large cavities. This represents the maximum reduction expected under these assumptions. In reality, only a fraction of the displaced gas is likely to be uplifted from the central region, while higher-entropy gas can flow inward on dynamical timescales comparable to the evolution of the cavities. The true density enhancement and reduction in cooling time are therefore likely to be smaller. For the simple linear uplift model adopted here, a shorter cooling time would proportionally reduce the distance over which the gas must be lifted before it can cool. Thus, the lifting radii derived in Fig.~\ref{fig:rad-vel} using the present day cooling times may be overestimated by up to a factor of $\sim2$. The actual correction depends on the efficiency with which cavities uplift central low-entropy gas, the subsequent inflow of higher-entropy material, and the degree of mixing, thermodynamic redistribution and the timing between successive AGN outbursts.

\begin{table*}
    \centering
    \begin{tabular}{lccccc}
        \toprule
        BCG & $t_{\rm cool}(R=10$ kpc) & $\dot{M}_{\rm Xray}$ ($R<$10 kpc) & $R_{\rm cool} \, (t_{\rm cool} = 1$ Gyr) & $L_{\rm Xray} \, (R<R_{\rm cool}$) & $\dot{M}_{\rm Hidden}$ \\
         &  (10$^8$ yr) & ($M_\odot$ yr$^{-1}$) & (kpc) & (erg s$^{-1}$) & ($M_\odot$ yr$^{-1}$) \\
        \midrule
                A85  &  $4.4\pm0.1$  &  21$^{+6}_{-6}$  &  17.1 & $(2.0\pm0.1)\times10^{43}$ &  25\\
                A133  &  $6.6\pm0.3$  &  24$^ {+1}_{-1}$  &  18.1  & $(25.7\pm0.1)\times10^{42}$ &  \\
                A262  &  $9.1\pm1.6$  &  3$^{+1}_{-1}$  &  12.3  & $(18.1\pm0.3)\times10^{41}$ &  6.6\\
                A1664  &  $3.9^{+2.2}_{-1.8}$  &  33$^{+18}_{-19}$  &  22.0  & $(3.1\pm0.2)\times10^{43}$  &  145 \\
                A1795  &  $8.6^{+4.7}_{-4.0}$  &  12$^{+6}_{-7}$  &  14.1  & $(1.5\pm0.1)\times10^{43}$  &  $>$21.6 \\
                A1835  &  $3.2^{+0.7}_{-0.6}$  &  111$^{+25}_{-27}$  &  40.7  &  $(7.6\pm0.1)\times10^{44}$  &  70 \\
                A2029  &  $7.7^{+2.0}_{-1.6}$  &  21$^{+4}_{-6}$  &  19.2  & $(7.3\pm0.3)\times10^{43}$ &  \\
                A2052  &  $5.4\pm0.1$  &  61$^{+1}_{-1}$  &  21.4  & $(30.8\pm0.1)\times10^{42}$  &  13.4 \\
                A2199  &  $6.8\pm0.7$  &  12$^{+1}_{-1}$  &  12.9  & $(7.8\pm0.2)\times10^{42}$  & 5.6 \\
                A2597  &  $4.2^{+0.5}_{-0.4}$  &  37$^{+4}_{-4}$  &  28.3  & $(8.2\pm0.1)\times10^{43}$  &  67 \\
                A2626  &  $12^{+5.6}_{-4.1}$  &  4$^{+1}_{-2}$  &  8.5  & $(1.3\pm0.2)\times10^{42}$  &   \\
                AS1101  &  $6.6\pm0.7$  &  24$^{+3}_{-3}$  &  22.9  & $(29.8\pm0.2)\times10^{42}$ & 10.7 \\
                Hydra-A  &  $8.9^{+1.8}_{-1.5}$  &  11$^{+2}_{-2}$  &  10.7  & $(9.1\pm0.2)\times10^{42}$ &  17 \\
                IC1262  &  $1.4^{+1.2}_{-1.1}$  &  2$^{+2}_{-1}$  &  3.1  & $(1.0\pm0.2)\times10^{41}$ &  \\
                MACS1347-11  &  $2.5\pm0.3$  &  386$^{+38}_{-44}$  &  40.4   &  $(23.3\pm0.3)\times10^{44}$ &  \\
                MACS1423+24  &  $3.1^{+0.3}_{-0.3}$  &  385$^{+34}_{-36}$  &  38.3   &  $(14.6\pm0.2)\times10^{44}$  &  \\
                MS0735+7421  &  $6.2^{+1.2}_{-1.3}$  &  42$^{+10}_{-11}$  &  25.8 & $(1.3\pm0.3)\times10^{44}$  &  \\
                NGC5044  &  $4.6\pm0.4$  &  7$^{+1}_{-1}$  &  18.9 & $(33.1\pm0.3)\times10^{41}$ &  21.6 \\
                PKS0745-191  &  $3.1\pm0.3$  &  87$^{+10}_{-11}$  &  38.1  &  $(46.9\pm0.4)\times10^{43}$  &   \\
                Phoenix  &  $2.0^{+0.4}_{-0.3}$  &  553$^{+84}_{-105}$  &  53.1  &  $(21.8\pm0.3)\times10^{45}$  & $>$800  \\
                RXJ1504.1-0248  &  $2.8\pm0.5$  &  162$^{+27}_{-22}$  &  46.0  & $(17.0\pm0.2)\times10^{44}$  & 757 \\
                Zw2701  &  $11^{+1.9}_{-1.8}$  &  28$^{+5}_{-6}$  &  21.8  &  $(10.4\pm0.5)\times10^{42}$  &   \\
                Zw7160  &  $4.8^{+1.8}_{-1.1}$  &  57$^{+15}_{-18}$  &  35.2  & $(3.5\pm0.1)\times10^{44}$  & \\
                ZwCl3146  &  $2.8\pm0.5$  &  197$^{+37}_{-40}$  &  33.6  &  $(4.9\pm0.1)\times10^{44}$  &  895 \\
        \bottomrule
    \end{tabular}
    \caption{X-ray cooling properties of the clusters in our sample derived from the deprojected profiles. The table includes the cooling time and cooling rate within 10 kpc, the cooling radius defined by $t_{\rm cool} = 1$ Gyr, and the corresponding X-ray cooling luminosity. We also include the total cooling rates (unabsorbed plus absorbed or hidden cooling) reported by \citet{fabian25}.
    }
    \label{tab:mcool}
\end{table*}

\subsection{The structure and morphology of the cooling and cold gas}
\label{cloud_shapes}

\begin{figure}
    \centering
    \includegraphics[width=0.48\textwidth]{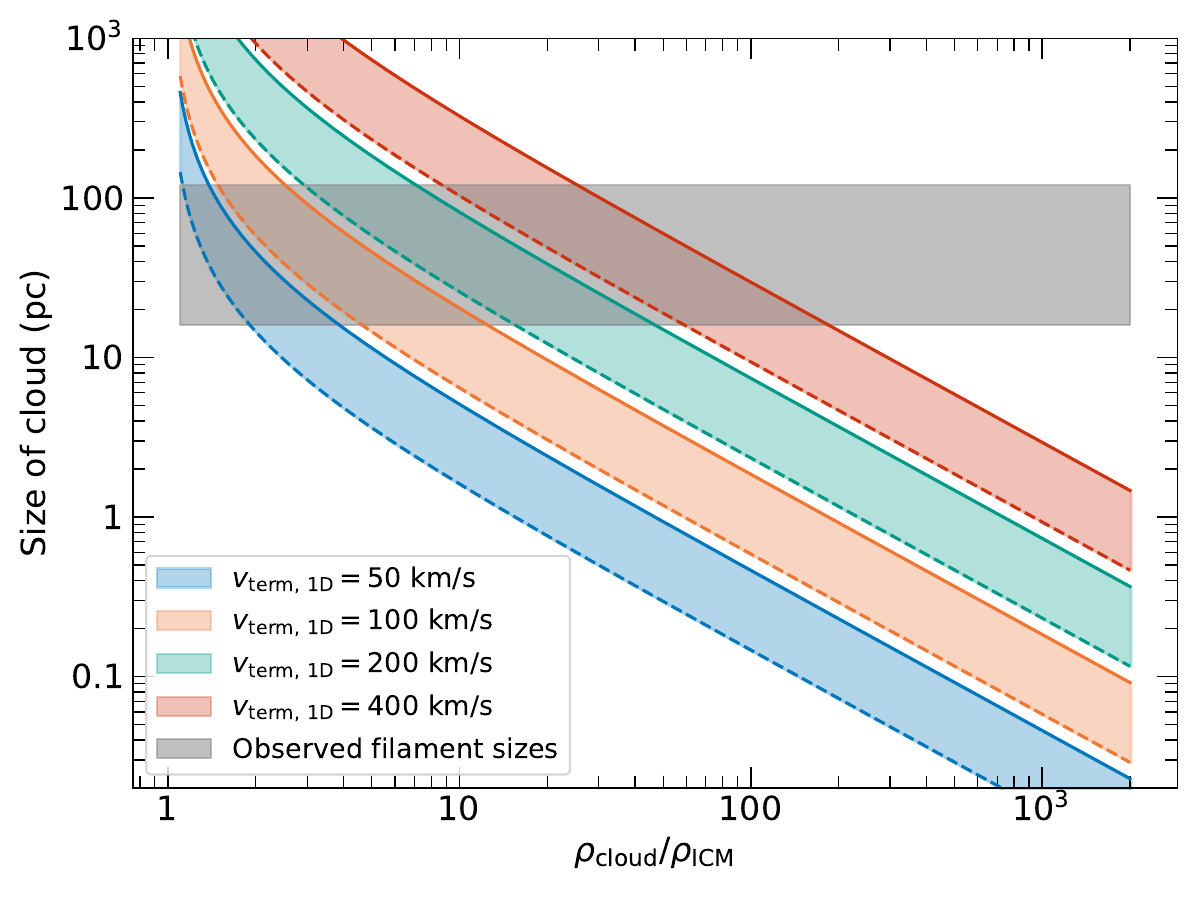}
    \caption{Estimated size of cooling clouds as a function of their density contrast with the surrounding ICM, for line-of-sight terminal velocities of 50, 100, 200, and 400 km s$^{-1}$. For each velocity, the shaded band spans the range of cloud sizes between a cluster-centric radius of 1 kpc (dashed boundary) and 10 kpc (solid boundary). The horizontal grey band marks the 16--120 pc widths of H$\alpha$ filaments observed in nearby central galaxies \citep{tamhane26}.}
    \label{fig:clsize}
\end{figure}

The precise size, mass and morphology of cooling clouds is not known. However, these parameters can be constrained if we assume that the clouds are overdense compared to the surrounding medium and are falling at their terminal velocities.
In this case, the force of gravity ($F_{\rm grav}$) is countered by the buoyancy ($F_{\rm buoy}$) and the drag force ($F_{\rm drag}$). Therefore,
\begin{equation}
    F_{\rm grav} = F_{\rm buoy} + F_{\rm drag},
\end{equation}
which becomes,
\begin{equation}
    \rho_{\rm cloud}\, V\, g = \rho_{\rm ICM}\, V\, g + \frac{1}{2} C_d\, \rho_{\rm ICM}\, A\, v_{\rm term}^2,
\end{equation}
where $g$ is the gravitational acceleration, $\rho_{\rm cloud}$ and $\rho_{\rm ICM}$ are the densities of the cloud and the surrounding medium respectively, $V$ is the volume of the cloud, $A$ is the cross sectional area of the cloud and $C_d$ is the drag coefficient. Rearranging the terms and solving for $v_{\rm term}$ gives
\begin{equation}
    v_{\rm term} \sim \sqrt{g \frac{V}{A}\frac{2}{C_d} \Biggl( \frac{\rho_{\rm cloud}}{\rho_{\rm ICM}}-1\Biggr)},
    \label{eqn:vterm}
\end{equation}

For simplicity, we assumed that the cooling gas clouds are spherical with a drag coefficient of $C_d \sim$0.47. In that case, $V/A = 2L/3$, where $L$ is the diameter of the cloud. Thus,
\begin{equation}
    v_{\rm term} \sim \sqrt{\frac{4}{3} \frac{gL}{C_d} \Biggl( \frac{\rho_{\rm cloud}}{\rho_{\rm ICM}}-1\Biggr)}.
    \label{eqn:vterm}
\end{equation}

We estimated the mean gravitational acceleration at radii of 1 and 10 kpc for each cluster using the potential profiles of \citet{hogan17b}, and averaged these over the sample to obtain representative values used in Figure~\ref{fig:clsize}. At these radii, the mean ICM number density spans $0.12$--$0.08$~cm$^{-3}$. Since the observed velocities are line-of-sight projections of the true 3D velocities, we correct for the random orientation of cloud motions by a factor of $\sqrt{3}$, such that $v_{\rm term,1D} = v_{\rm term}/\sqrt{3}$.

Figure~\ref{fig:clsize} shows the inferred cloud sizes as a function of density contrast $\rho_{\rm cloud}/\rho_{\rm ICM}$ for four representative line-of-sight terminal velocities, with shaded bands indicating the range between 1 and 10~kpc used to estimate the gravitational acceleration to estimate the terminal velocities at that radius. At a density contrast of $\sim$100, a cloud moving at 200~km~s$^{-1}$ would have a characteristic size of 5--15~pc across this radial range, somewhat smaller than the observed widths of cold gas filaments in BCGs \citep[e.g.,][]{fabian08,fabian16,anderson18,tamhane26}. For the lower velocities more typical at 10~kpc ($\lesssim$100~km~s$^{-1}$), the implied cloud sizes decrease further to a few parsecs or less at the same density contrast. Conversely, for clouds with sizes comparable to the observed filament widths of 70--120~pc, the required density contrast is $\sim$10--50 at 10~kpc. 

For denser cooling clouds with density contrasts of $\sim$1000, the inferred cloud sizes become extremely small, ranging from a few parsecs down to sub-parsec scales. This is broadly consistent with simulations suggesting that cooling gas in cluster cores may fragment into a mist-like population of compact cloudlets confined by the surrounding ICM pressure, turbulence, and magnetic fields \citep{mccourt12,mccourt18,farber23}. At 10~kpc, a density contrast of $\sim$1000 corresponds to a cloud hydrogen number density of $\sim$80~cm$^{-3}$, given the mean ICM density of $\sim$0.08~cm$^{-3}$ at that radius. This is comparable to the electron densities of $\sim$10$-$100~cm$^{-3}$ inferred from optical [\ion{S}{2}] line ratios in warm ionized filaments in cool-core clusters \citep[e.g.,][]{hamer16}, broadly consistent with density contrasts of order $\sim$100$-$1000 in these structures. We note that [\ion{S}{2}] traces only the warm ionized component of the multiphase filaments, while the cold molecular and atomic phases are likely denser still.

These estimates are approximate, as the calculation assumes spherical clouds with a drag coefficient of $C_d = 0.47$, whereas the observed structures are elongated filaments with large aspect ratios. Furthermore, we assume that the filaments move at their terminal velocities, which may not hold at all radii. Magnetic fields amplified during gas condensation may also help stabilize elongated structures against disruption, maintain the narrow coherent morphologies observed in BCG filaments, and reduce the effective infall velocities of cold clouds \citep[e.g.,][]{tamhane26,voit26,fournier26}. Accounting for these effects would increase the inferred cloud sizes.

\subsection{How much gas is expected to have cooled?}
\label{sec:cooling}

The radiative cooling rate can be estimated as the ratio of the gas mass to its cooling time within a given radius:

\begin{equation}
\dot{M}_{\rm Xray} = \frac{M_{\rm gas}(r < r_c)}{t_{\rm cool}},
\end{equation}

where the cooling time $t_{\rm cool}$ is calculated using equation~\ref{eqn:tcool} and $r_c$ is the cooling radius. Since we are interested in the uplift of low-entropy gas from the central 10 kpc, we estimate radiative cooling rates using a fixed $r_c = 10$ kpc for all clusters. The total hot gas mass within 10 kpc is derived from the deprojected X-ray gas mass profiles. Table~\ref{tab:mcool} shows the cooling times, and cooling rates for clusters in this sample.

\begin{figure*}
    \centering
    \includegraphics[width=0.8\textwidth]{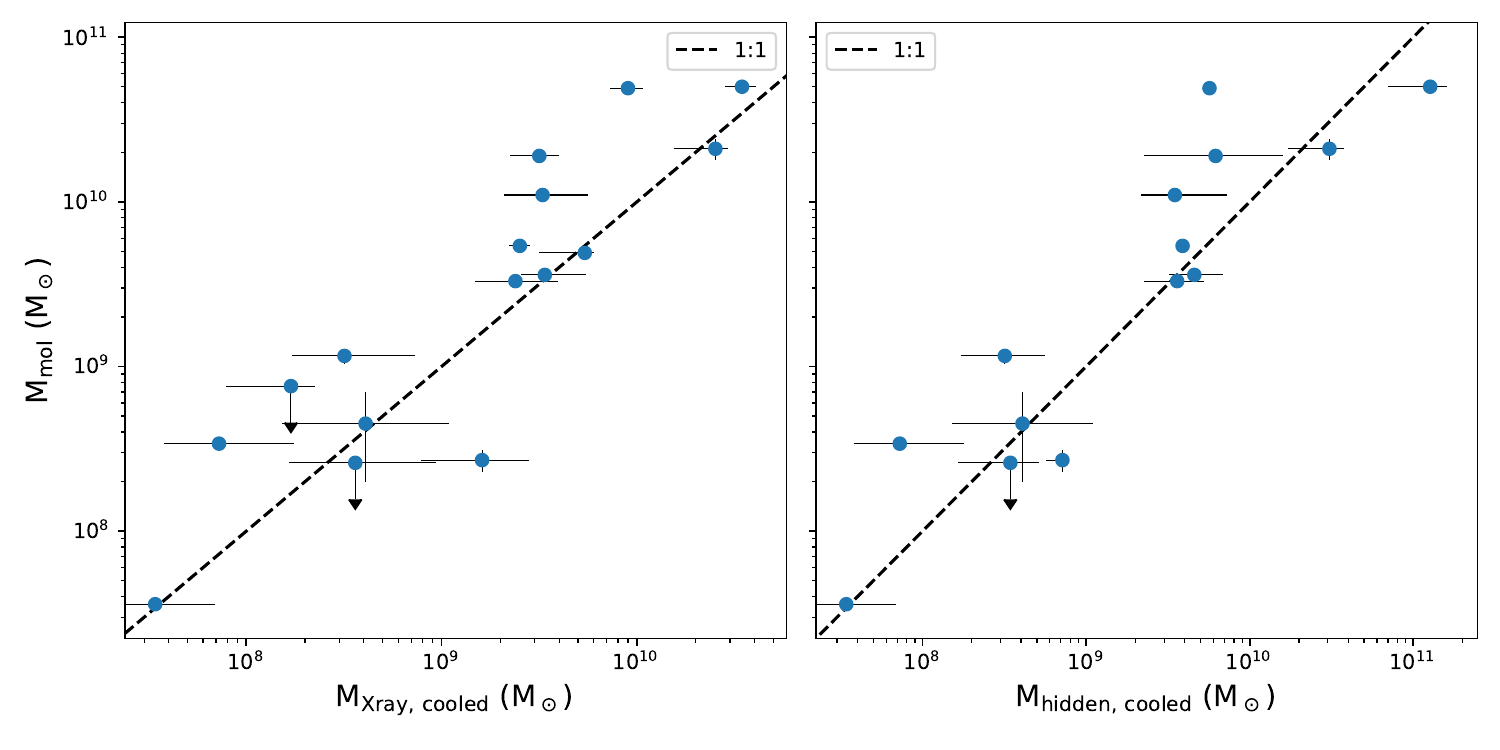}
    \caption{Comparison between the amount of gas expected to have cooled and the observed molecular gas mass in BCGs. The left panel shows the expected cooled mass based on radiative X-ray cooling rates, while the right panel shows the expected cooled mass using the hidden cooling rates. The dashed grey line in both panels shows the 1:1 relation.}
    \label{fig:mcooled}
\end{figure*}

To estimate the mass of gas that could have cooled over the lifetimes of the currently observed cavities, which we denote as $M_{\rm Xray,\,cooled}$, we multiplied the inferred X-ray cooling rates by the estimated cavity ages. This assumes that the cooling rate remained approximately constant over the cavity lifetime and that the cooling gas could ultimately contribute to the cold molecular reservoir. We ignored large cavities in Hydra-A and MS0735-7421. We additionally limit the maximum cooled gas mass to $0.5\,M_{\rm disp}$, motivated by analytical models which suggest that the mass of gas lifted by a spherical bubble is approximately half of the mass displaced by the bubble \citep{pope10}. This therefore provides an approximate upper limit on the amount of gas lifted by the currently observed cavities that could subsequently cool and contribute to the observed molecular gas reservoir. This estimate assumes a 100\% cooling efficiency, such that all uplifted gas eventually cools.

In the left panel of Figure~\ref{fig:mcooled}, we compare these expected cooled gas masses with the observed total molecular gas masses. In 4 out of 12 systems, $M_{\rm mol}/M_{\rm Xray,\,cooled}<1$. However, only A2052 and A2199 contain clearly less molecular gas than estimated from cooling within cavity lifetimes, with ratios of 0.17 and $<0.72$, respectively. Phoenix and PKS0745 have ratios of $\sim0.9$, consistent with unity within the uncertainties. Three additional systems, A2597, A85, and NGC~5044, also have $M_{\rm mol}/M_{\rm Xray,\,cooled}\approx1$ within the uncertainties. Thus, five systems have ratios within approximately $1\pm0.2$. In the remaining systems, the observed molecular gas mass is $\sim1.5$--6 times larger than estimated.

The inferred cooling rates may, however, underestimate the true amount of gas cooling below $\sim2$ keV. The soft X-ray luminosities of many cool-core clusters are lower than expected from classical cooling-flow models \citep{peterson01,tamura01a}, and cold gas associated with the multiphase medium may absorb some of this emission or promote additional cooling through mixing \citep{fabian02,soker04,fabian11}. \citet{fabian23,fabian23b} proposed a hidden cooling-flow model in which soft X-ray emission is absorbed by intervening cold gas and used XMM-Newton RGS spectra to infer corrected cooling rates for several cool-core systems. Adopting these hidden cooling rates marginally increases the expected cooled gas masses but does not significantly change the comparison with the observed molecular gas in most systems. The main exception is ZwCl3146, for which the estimated cooled gas mass exceeds the observed molecular gas mass, as shown in the right panel of Figure~\ref{fig:mcooled}. Thus, hidden cooling has only a modest effect on the overall mass comparison.

The comparisons above use the total observed molecular gas mass in each BCG. However, a substantial fraction of this gas, ranging from 10\% to 70\% with an average of $\sim$50\%, lies in extended filaments that are more directly associated with the observed cavities \citep{russell19,tamhane22}. Restricting the comparison to this filamentary molecular component would therefore bring the estimated cooled gas mass and observed molecular gas mass into closer agreement in some systems. However, several systems would still contain more molecular gas than expected from cooling associated with the currently observed cavities.

These results indicate that the present day molecular reservoir cannot, in general, be attributed solely to cooling associated with the currently observed cavities in all systems. Even in systems where $M_{\rm mol}$ is comparable to $M_{\rm Xray,\,cooled}$, this agreement depends on the assumed lifting and cooling efficiencies. If cavities lift less than $50$\% of the displaced gas mass, or if only a fraction of the uplifted gas subsequently cools, the estimated cooled mass would decrease and the ratio $M_{\rm mol}/M_{\rm Xray,,cooled}$ would increase. The molecular gas may therefore have accumulated over several feedback cycles. Radio-mode feedback in cool-core clusters has a high duty cycle, with cavities inferred to be present for at least $\sim60$--70\% of the time and potentially nearly continuously \citep{birzan12}. This raises the possibility that some fraction of the cold gas formed during earlier feedback episodes survives subsequent AGN outbursts, while later cycles lift additional low-entropy gas and promote further condensation. The present day molecular reservoir may therefore represent the cumulative outcome of repeated uplift and cooling over several feedback cycles rather than the product of a single outburst.

\subsection{Multiphase emission and the X-ray cooling luminosity}

\begin{figure*}
    \centering
    \includegraphics[width=\textwidth]{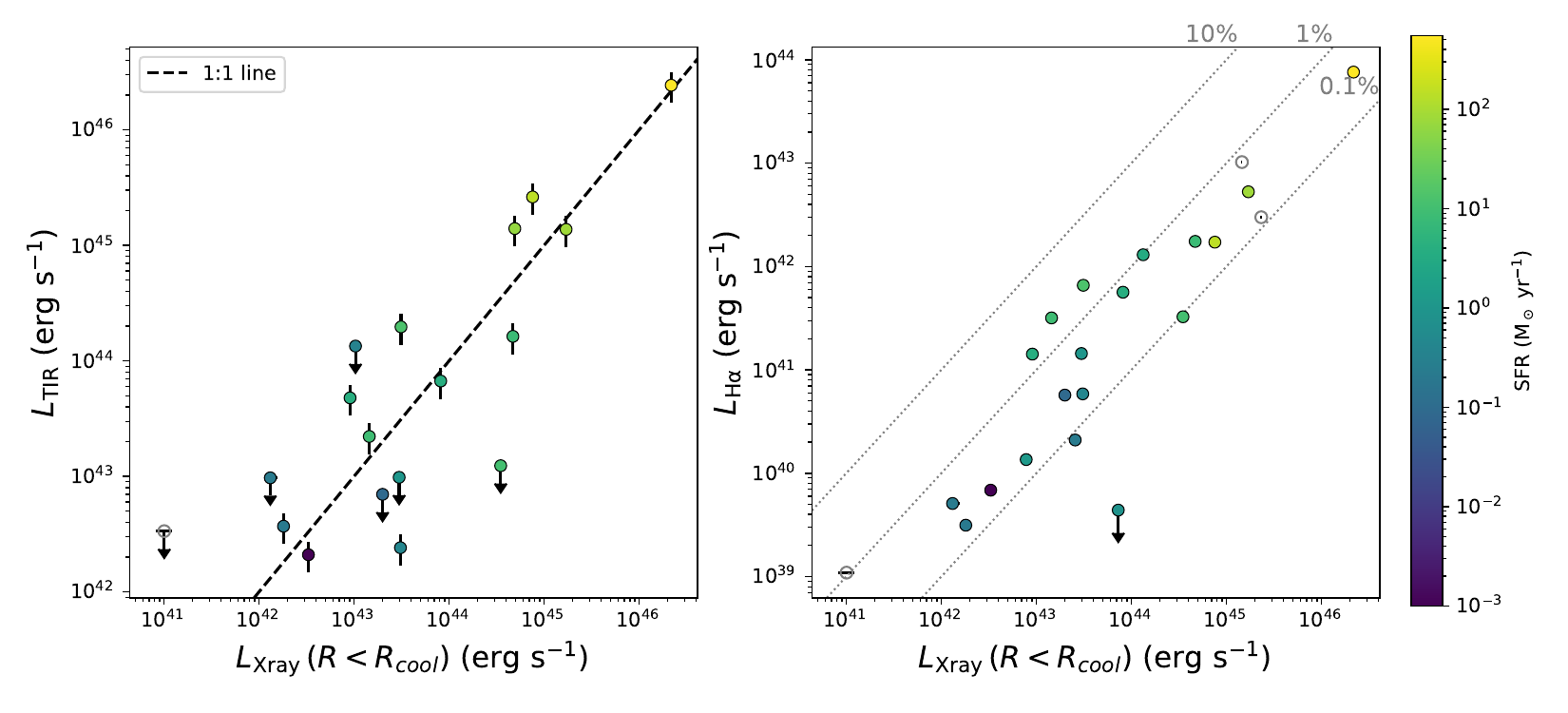}
    \caption{Total infrared luminosities ($L_{\rm TIR}$) and H$\alpha$ luminosities ($L_{\rm H\alpha}$) are plotted against X-ray cooling luminosities ($L_{\rm Xray}$) in the {\em left} and {\em right} panels, respectively. Points are colored by the SFR in the logarithmic scale as shown by the colorbar. Clusters lacking SFR measurement are shown as open grey circles. The X-ray luminosities are measured within a cooling radius $R_{\rm cool}$ defined by the radius at which gas cooling time $t_{\rm cool} = 1$ Gyr. The black dashed line marks the 1:1 relation in the {\em left panel}. The dotted grey lines in the {\em right panel} indicate the constant H$\alpha$-to-X-ray cooling luminosity fractions, as labeled.
    }
    \label{fig:McoolComp}
\end{figure*}

The cooling of the ICM is expected to produce emission across multiple phases as gas loses thermal energy. In addition to X-ray emission in the hot phase, some of this energy emerges at optical and infrared wavelengths as the gas cools to lower temperatures. BCGs commonly contain dust and multiphase gas \citep[e.g.,][]{edge10,donahue11}, providing a potential mechanism for absorbing and re-radiating some the cooling luminosity. We therefore compare X-ray luminosity with optical H$\alpha$ and total IR luminosity to assess whether luminosities can account for a significant fraction of the total cooling luminosity.

We estimate the X-ray cooling luminosity ($L_{\rm Xray}$) within a cooling radius ($R_c$) defined by $t_{\rm cool} = 1$ Gyr from deprojected profiles. These values are provided in Table~\ref{tab:mcool}. We have not subtracted the contribution from X-ray AGN in six sources in which they are detected, therefore, luminosities in those clusters are overestimated by up to $\lesssim 10$\% except in Phoenix, where the AGN contribution could be up to 50\% \citep{mcdonald19}.

We first compare $L_{\rm Xray}$ with the observed H$\alpha$ luminosities provided in Table~\ref{tab:cluster_properties}. Interpreting the H$\alpha$ emission entirely as a consequence of cooling would provide an estimate of the fraction of the cooling luminosity emitted in this line. However, H$\alpha$ emission in BCGs can also be powered by other mechanisms, including star formation and photoionization by the active galactic nucleus. In particular, several BCGs in our sample host ongoing star formation. We therefore regard the ratio $L_{\rm H\alpha}/L_{\rm Xray}$ as an upper limit to the fraction of the cooling luminosity that can be attributed to H$\alpha$ emission rather than as a direct measurement of the cooling fraction. The right panel of Figure~\ref{fig:McoolComp} shows the comparison between the two luminosities. The H$\alpha$ luminosities are typically $\lesssim 1$\% of the cooling luminosity emitted in the X-ray. We find an approximately linear relationship between $L_{\rm H\alpha}$ and $L_{\rm Xray}$, consistent with previously reported connection between H$\alpha$ filaments and the cooling ICM \citep[e.g.,][]{mcdonald10,olivares25}. Interestingly, we do not find a significant offset in the relationship with SFR. This may suggest that although star formation contributes to the H$\alpha$ emission, it is mainly associated with ICM cooling.

Dust provides another possible channel for the reprocessing of cooling radiation. Ultraviolet and optical emission produced during the interaction and mixing of hot and cold gas can be absorbed by dust and subsequently re-radiated in the infrared. In addition, dust mixed with the hot gas can enhance cooling by increasing the radiative cooling efficiency through collisional interactions and by providing additional cooling channels \citep{silk74}. Dust has indeed been observed in many cool-core clusters \citep{edge10,donahue11}. We therefore compare the observed total infrared luminosity, $L_{\rm TIR}$, with $L_{\rm Xray}$.

We use the method described in \citet{dale14} to estimate the total infrared luminosity. The relation in \citet{dale14} is expressed in terms of the monochromatic luminosities at 24, 70, and 160~$\mu$m, where $L_{24}$, $L_{70}$, and $L_{160}$ denote $\nu L_\nu$ at the corresponding wavelengths as follows:
\[
L_{\rm TIR}=1.548\,L_{24}+0.767\,L_{70}+1.285\,L_{160},
\]
which corresponds to an assumed AGN contribution of 0\%. Following \citet{donahue11}, we assume $\nu L_\nu(70\,\mu{\rm m})\simeq\nu L_\nu(160\,\mu{\rm m})$ when 160~$\mu$m measurements are unavailable. For the five clusters in our sample with detected X-ray AGN and IR data (A2052, Hydra-A, NGC~5044, Phoenix, and PKS~0745), we instead assume an AGN contribution of 25\% and use the corresponding relation from \citet{dale14}:
\[
L_{\rm TIR}=1.574\,L_{24}+0.763\,L_{70}+1.352\,L_{160}.
\]
We then adopt 75\% of this luminosity as the non-AGN infrared emission. These values are reported in Table~\ref{tab:cluster_properties} and used in our analysis. The $Spitzer$-derived infrared flux densities for clusters in our sample were obtained from NED\footnote{\url{https://ned.ipac.caltech.edu/}}. In Phoenix and RXCJ1504, we adopted 70 and 160~$\mu$m flux densities from Herschel/PACS observations. Infrared measurements were available for 18 clusters in our sample, with upper limits for six clusters. For the Phoenix cluster, we also adopted the WISE 22~$\mu$m flux density in place of the $Spitzer$ 24~$\mu$m flux density because $Spitzer$ measurements were unavailable. The flux densities were converted to luminosities using
\[
\nu L_\nu=\frac{4\pi D_L^2F_\nu\nu}{1+z},
\]
where $D_L$ is the luminosity distance and $z$ is the redshift.

We next compare the X-ray cooling luminosity with the total infrared luminosity, $L_{\rm TIR}$. The left panel of Figure~\ref{fig:McoolComp} shows an approximately linear relation between $L_{\rm TIR}$ and $L_{\rm Xray}$, with systems having larger X-ray cooling luminosities generally exhibiting larger infrared luminosities. However, the infrared luminosity is not consistently smaller than the X-ray cooling luminosity. Some systems have $L_{\rm TIR}<L_{\rm Xray}$, while others have $L_{\rm TIR}>L_{\rm Xray}$, with a mean ratio $\langle L_{\rm TIR}/L_{\rm Xray}\rangle>1$. Thus, the observed infrared luminosity cannot be interpreted as a direct measure of the fraction of the X-ray cooling luminosity that is re-radiated by dust. In addition, our $L_{\rm TIR}$ estimates are based on empirical conversions from a limited number of infrared bands \citep{dale14}, and therefore carry systematic uncertainties associated with the assumed infrared spectral energy distribution. More detailed SED modeling with mid- and far-infrared data would be required to separate the contributions from star formation, AGN-heated dust, and emission potentially associated with cooling.

Nevertheless, the approximately linear relation between $L_{\rm TIR}$ and $L_{\rm Xray}$ suggests that the infrared-emitting material is closely connected to the cooling atmosphere. Star formation likely contributes substantially to $L_{\rm TIR}$ in many systems, but this does not necessarily make the infrared emission unrelated to cooling, since the star formation itself is fueled by the cooled gas reservoir. The absence of a clear systematic offset from the relation with increasing SFR is therefore consistent with $L_{\rm TIR}$ tracing the broader cooling cycle, although it cannot be interpreted as a direct measure of the instantaneous cooling luminosity. A particularly useful case is NGC~5044, where UV observations indicate negligible ongoing star formation and a far-infrared SED fit with a modified blackbody yields a slightly lower $L_{\rm TIR}$ than the empirical estimate adopted here \citep{tamhane26}. Using our empirical estimate gives $L_{\rm TIR}/L_{\rm Xray}\simeq0.8$, while adopting the SED-based luminosity and allowing for a similar 25\% AGN contribution reduces the ratio to $\sim0.5$. This suggests that, at least in systems with little ongoing star formation, a substantial fraction of the cooling luminosity may emerge in the infrared, although a more precise fraction remains uncertain because of the decomposition due to different heating sources.

\subsection{A note on Abell 2029 and Zw7160}

Abell 2029 and Zw7160 provide an interesting contrast because neither system hosts clear X-ray cavities, yet their cold gas contents are very different. Abell 2029 has low central entropy and a short cooling time and hosts the extended radio source PKS~1508+059, reaching $\sim40$ kpc, but has no detectable molecular gas \citep{martz20}. The core also exhibits prominent sloshing, and the radio lobes appear to interact with the sloshing atmosphere \citep{pm13}. Therefore, the lack of detectable cavities possibly imply that cavities formed during earlier stages of the radio outburst may have been distorted or disrupted by pre-existing sloshing motions or hydrodynamical instabilities \citep{pm13,xrism_a2029}.

Zw7160 (MS1455.0+2232), however, presents a different scenario. It also lacks prominent X-ray cavities, has low central gas cooling time, exhibits strong large-scale sloshing and its central radio source is relatively compact \citep{riseley22}, yet it contains one of the largest molecular gas reservoirs in our sample, $M_{\rm mol}\sim6\times10^{10}$ M$_\odot$. \citet{mazzotta01} proposed that its large cold gas mass may have accumulated by an unusually long-lived cooling flow, with a strongly cooling atmosphere established prior to the present dynamical configuration, rather than condensation driven by the currently observed sloshing itself. The sloshing may then redistribute the low-entropy gas without destroying the pre-existing cold gas \citep{mazzotta08}. The comparison between A2029 and Zw7160 therefore shows that the absence of presently detectable cavities does not always determine the cold gas content. Instead, the formation of a molecular reservoir likely depends on the dynamical and thermodynamic history of the core and on how effectively different forms of gas motion transport low-entropy material into conditions favorable for condensation.

\section{Limitations and Caveats}

Several factors affect the interpretation of our results, stemming from the assumptions underlying our calculations and other physical channels not included.

The molecular gas masses used in this study were estimated using the Galactic CO-to-H$_2$ conversion factor ($X_{\rm CO}$), except in Phoenix where five times lower value is used \citep[see for e.g.,][]{russell17b}. However, if $X_{\rm CO}$ is systematically lower in BCGs compared to its Galactic value, the inferred molecular gas masses will also be proportionally lower, which would reduce or in some systems potentially eliminate the discrepancy between displaced and/or potentially cooled gas mass and present day molecular gas mass. This is an important systematic to revisit once a robust calibration of $X_{\rm CO}$ in a representative sample of central galaxies of cool-core clusters becomes available.

In addition, uplift by buoyant cavities is unlikely to be the only mechanism capable of redistributing low-entropy gas. Sloshing and mergers can also displace or transport low-entropy gas to larger radii and, under suitable thermodynamic conditions, may promote condensation \citep{olivares23}. Such bulk motions can also distort, disrupt, or erase cavities produced by previous AGN outbursts, as may occur in systems such as Abell 2029. Therefore, the properties of the currently detected cavities may not provide a complete feedback history responsible for the present day molecular gas reservoir. The timescales for which X-ray cavities and multiphase filaments are visible may differ, such that cavities associated with earlier cooling episodes may have expanded, dissipated, been disrupted by ICM motions, or become difficult to detect while the condensed gas persists. These effects may contribute to the scatter observed in the $M_{\rm disp}$--$M_{\rm mol}$ relation discussed in Section~\ref{sec:lifting}.

Beyond these assumptions, our calculations do not account for several additional processes that could contribute to the cold gas budget. In-situ cooling of the ambient ICM, triggered by thermal instabilities associated with AGN feedback, can produce cold gas and contribute to the total molecular gas reservoir \citep[e.g.,][]{voit17,gaspari17,jennings23,sotira26,fournier26}. In practice, both lifting-driven and in-situ cooling channels likely operate simultaneously.

Cavity-driven uplift and cooling may nonetheless play an important role, because cavities can lift substantial amounts of low-entropy gas from the cluster core, where the low-entropy gas reservoir is largest and where dust originating from the BCG may be mixed into the ambient gas. The presence of dust can enhance cooling efficiency and its survival in cluster cores may play a key role in enabling the formation of molecular gas \citep{temi26,tamhane26}. In contrast, in-situ cooling triggered by shocks, turbulence, or compression around cavity rims is expected to occur in more localized regions and may therefore contribute a smaller fraction of the total cold gas mass \citep[e.g.,][]{sotira26}. Nevertheless, such localized cooling could enhance or replenish existing cold gas structures over time.

A related channel we do not model is the turbulence driven mixing in the cold/hot gas interfaces. Numerical simulations indicate that mixing between phases facilitated by turbulence, can accelerate cooling by transporting low-entropy gas into the hotter medium \citep{mohapatra23}. More specifically, cold gas may form at cavity rims and filament boundaries via turbulent radiative mixing layers, in which turbulent mixing drives intermediate temperature gas to cool radiatively. This process can potentially grow existing filaments by adding mass from hot to the cold phase than simple ram-pressure or entrainment/wake estimates \citep[e.g.,][]{fielding22,tan24,cammelli26}. These processes are difficult to isolate observationally, but may help explain the scatter in the molecular gas masses across the sample.

Finally, extreme systems such as the Phoenix cluster exhibit extremely short central cooling times of only a few $\times10^7$ years. Although Phoenix is likely an exceptional case, it raises the possibility that clusters may undergo brief phases of rapid cooling with minimal uplift. If such episodes occur intermittently, clusters observed at different stages of the AGN feedback cycle could display a wide range of cold gas masses and filament morphologies.

\section{Conclusions}

We have used Chandra observations of 24 cool-core clusters together with measurements of X-ray cavities and multiphase gas to test whether cavity-driven uplift can account for the formation of molecular gas in BCGs. Our main conclusions are summarized below:

\begin{enumerate}

\item Almost all clusters contain sufficient hot gas within the central $\sim$10 kpc region to supply the observed molecular gas reservoir. The supply of low-entropy gas is therefore generally not the limiting requirement for gas uplift and condensation.

\item The maximum gas mass displaced by the observed cavities is comparable to the molecular gas mass in many systems. Across the full sample, the median $M_{\rm mol}/M_{\rm disp}$ is $\sim0.35$, while systems satisfying $M_{\rm mol}<M_{\rm disp}$ require, on average, only $\sim27$\% of the maximum displaced gas mass to be comparable to the total molecular gas reservoir. Restricting the comparison to the filamentary molecular component reduces the requirement to $\sim14$\%.

\item The molecular and displaced gas masses scale approximately linearly, although with substantial intrinsic scatter. This suggests that while uplift plays an important role, other factors, such as dynamical disturbances, and ICM--cavity coupling and other cooling channels affect the final molecular reservoir. In particular, some systems with very large or spatially extended cavities lie well below the best-fit relation, suggesting that the most powerful cavity systems do not necessarily couple efficiently to the central low-entropy gas and may therefore contribute less effectively to cold gas formation.

\item For gas with short cooling times ($\sim10^8$ yr), the observed uplift velocities (150--400 km s$^{-1}$) are sufficient to transport gas to typical H$\alpha$ filament radii of 10--30 kpc. If the gas is transported in circular motions driven by turbulent eddies, the required radii of 30--60 kpc are broadly consistent with observed line-of-sight velocity dispersions and inner cooling times. Gas with longer cooling times ($\sim10^9$ yr) require much larger uplift radii ($\gtrsim$60--200 kpc) and may cool during its return toward the cluster center.

\item Classical X-ray cooling over the lifetimes of the currently observed cavities could produce molecular gas masses comparable to the observed reservoirs in several systems under the assumed lifting and cooling efficiencies. However, many systems contain more molecular gas than expected from the currently observed cavities, suggesting that their molecular gas reservoirs may have accumulated over multiple AGN feedback cycles.

\item Both H$\alpha$ and infrared luminosities increase approximately linearly with the X-ray cooling luminosity, supporting a close connection between the multiphase gas and the cooling atmosphere. H$\alpha$ typically represents only $\sim1$\% of the X-ray cooling luminosity, while $L_{\rm TIR}$ can be comparable to or exceed $L_{\rm Xray}$. The infrared luminosity therefore cannot be used as a direct tracer of the instantaneous cooling luminosity, although it may still trace the broader cooling cycle through dust heating by star formation fueled by the cooled gas reservoir. However, in systems such as NGC~5044, where ongoing star formation is negligible, the large $L_{\rm TIR}/L_{\rm Xray}$ ratio suggests that a substantial fraction of the cooling luminosity may emerge in the infrared.

\end{enumerate}

Overall, our analysis shows that cavity-driven lifting can plausibly account for a substantial fraction of the molecular gas observed in cluster cores. In many systems, the mass of gas displaced by cavities is sufficient to produce the observed cold gas if only a modest fraction of the lifted gas condenses. However, the molecular gas mass often exceeds the amount expected to cool during the lifetime of a single cavity generation, suggesting that cold gas reservoirs likely accumulate over multiple AGN feedback cycles.

These results support a scenario in which cavity-driven lifting plays a central role in triggering gas condensation, while repeated AGN outbursts and additional cooling processes contribute to building the large molecular gas reservoirs observed in BCGs. The efficiency with which radio jets lift gas in cluster cores may also help explain why molecular gas flows in BCGs are substantially larger and more massive than those observed in other active galaxies. This highlights the importance of radio-mechanical feedback operating in hot atmospheres in regulating the baryon cycle in massive galaxies.

Despite this progress, uncertainties remain in our understanding of gas cooling and uplift. Accurate estimates of star formation rates, infrared luminosities, and the AGN contribution to the infrared budget are essential for disentangling the different cooling channels. High-resolution spectroscopy of intermediate-temperature gas and improved measurements of ICM turbulence will help constrain the efficiency of turbulent mixing and uplift-driven condensation. Infrared diagnostics of dust and improved measurements of the total molecular gas content, particularly with facilities such as the \textit{James Webb Space Telescope}, will provide a more complete picture of the cooling process. Together, these observations will help clarify the relative importance of different cooling channels and improve our understanding of the AGN feedback cycle.

\begin{acknowledgments}
We thank Greg Bryan for his insightful comments and suggestions. BRM acknowledges support from the Natural Sciences and Engineering Council of Canada and the Canadian Space Agency Space Science Enhancement Program. This research has made use of data obtained from the Chandra Data Archive and the Chandra Source Catalog, and software provided by the Chandra X-ray center (CXC) in the application packages CIAO and Sherpa.
\end{acknowledgments}

\begin{contribution}

PT analyzed the data and wrote the manuscript. BM supervised the research and contributed to the interpretation and discussion of the results. PN contributed to the interpretation and discussion of the results and provided detailed comments on the manuscript.


\end{contribution}

%
\facilities{CXO}

\software{astropy \citep{astropy13,astropy18,astropy22}
          }


\appendix

\section{Appendix information}

Table describing multiphase gas properties and star formation rates in BCGs.

\begin{table*}
\label{tab:cluster_properties}
\begin{tabular}{lccccc}
\toprule
Cluster  & Log H$\alpha$  & M$_{\rm mol}$ & $L_{\rm TIR}$ & SFR  & $\sigma_{\rm los}$ \\
& (erg s$^{-1}$) & (M$_\odot$) & ($\times10^{43}$ erg s$^{-1}$) & (M$_\odot$ yr$^{-1}$) & (km s$^{-1}$) \\
& (1) & (2) & (3) & (4) & (5) \\
\midrule
A85  & 40.756 $^g$  & (4.5$\pm$2.5)$\times$10$^8$$^g$  & $<1$ & 0.09 $^d$  &  145 $^c$ \\

A133  & 40.322 $^g$  & -  & - & 0.23 $^d$  & 132 $^c$ \\

A262  & 39.497 $^g$  & (3.4$\pm$0.1)$\times$10$^8$ $^h$  & 0.37$\pm$0.04 & 0.22 $^d$ & 110 $^f$ \\

A1664  & 41.818 $^g$  & (1.1$\pm$0.1)$\times$10$^{10}$ $^{i,h}$  & 19.7$\pm$1.0 & 13.18 $^d$  & 135 $^f$ \\

A1795  & 41.503 $^g$  & (3.3$\pm$0.2)$\times$10$^9$ $^{i,h}$  & 2.22$\pm$0.22 & 9.3 $^j$  & 151 $^f$ \\

A1835  & 42.235 $^g$  & (4.9$\pm$0.3)$\times$10$^{10}$ $^{i,h}$  & 263$\pm$30 & 150 $^b$  & 120 $^f$ \\

A2029  & $<$39.643 $^g$  & $<1.7\times$10$^9$ $^g$  & - & 0.87 $^d$  &  \\

A2052  & 40.768 $^g$  & (2.7$\pm$0.4)$\times$10$^8$ $^h$  & 0.3$\pm$0.01 & 0.44 $^d$  & 194 $^c$ \\

A2199  & 40.132 $^g$  & $<2.6\times$10$^8$ $^g$  & - & 1.23 $^d$  &  \\

A2597  & 41.751 $^g$  & (3.6$\pm$0.1)$\times$10$^9$ $^{i,h}$  & 6.7$\pm$0.3 & 3.98 $^d$  & 175 $^f$ \\

A2626  & 39.707 $^g$  & $<7.6\times$10$^8$ $^g$  & $<1$ & 0.23 $^d$  &  \\

AS1101  & 41.158 $^g$  & (1.2$\pm$0.1)$\times$10$^9$ $^i$  & $<1$ & 1.02 $^d$  & 158 $^f$ \\

Hydra-A  & 41.153 $^g$  & (5.4$\pm$0.2)$\times$10$^9$ $^h$  & 5.87$\pm$0.7 & 4.07 $^d$  &  142 $^f$ \\

IC1262  & 39.038 $^g$  & -  & $<0.34$ &    &  \\

MACS1347-11  & 42.477 $^g$  & $<6.8\times$10$^{10}$ $^g$  & - &    &  \\

MACS1423+24  & 43.010 $^g$  & -  & - &   &  \\

MS0735+7421  & 42.113 $^g$  & $<10^{9}$ $^g$  & - & 3.24 $^d$    &  \\

NGC5044  & 39.836 $^g$  & (3.6$\pm$0.3)$\times$10$^7$ $^{i,h}$  & 0.26$\pm$0.01 & $<0.001$ $^d$ & 196 $^k$ \\

PKS0745-191  & 42.243 $^g$  & (4.9$\pm$0.3)$\times$10$^9$ $^{i,h}$  & 20$\pm$1 & 8 $^b$  & 196 $^f$ \\

Phoenix  & 43.881 $^m$ & (2.1$\pm$0.3)$\times$10$^{10}$ $^{i,h}$  & 2990$\pm$41 & 550 $^e$  & 191 $^*$ $^i$ \\

RXJ1504.1-0248  & 42.723 $^g$ & (1.9$\pm$0.1)$\times$10$^{10}$ $^{i,h}$  & 138$\pm$4 & 85.11 $^d$  & 135 $^*$ $^i$ \\

Zw2701  & - & -  & $<13.4$ & 0.35 $^d$   &  \\

Zw7160  & 41.514 $^g$ & (5.8$\pm$2.5)$\times$10$^{10}$ $^g$  & $<1.24$ & 10.72 $^d$   &  \\

ZwCl3146  & 42.657 $^g$ & (5.0$\pm$0.5)$\times$10$^{10}$ $^l$  & 140$\pm$18 & 69.18 $^d$  & 150\*$^l$ \\
\bottomrule
\end{tabular}
\caption{Properties of molecular and H$\alpha$ gas masses and velocities for clusters in our sample. The column descriptions are provided below: (1) the log of the H$\alpha$ luminosity, (2) the total molecular gas mass, (3) the total infrared luminosity (4) the UV star formation rate, (5) the line of sight velocity dispersion of the warm gas. Values marked by * were estimated from the molecular gas velocity maps. References: $^a$: \citet{edge01}, 
$^b$: \citet{gingras25}, 
$^c$: \citet{hamer16}, 
$^d$: \citet{mcdonald18}, 
$^e$: \citet{mcdonald19}, 
$^f$: \citet{olivares19}, 
$^g$: \citet{pulido18}, 
$^h$: \citet{russell19}, 
$^i$: \citet{tamhane22}, 
$^j$: \citet{tamhane23}, 
$^k$: \citet{tamhane26}, 
$^l$: \citet{vantyghem21},
$^m$: \citet{mcdonald14b} .}
\end{table*}


\bibliography{sample701}{}

\begin{thebibliography}{}
\expandafter\ifx\csname natexlab\endcsname\relax\def\natexlab#1{#1}\fi
\providecommand{\url}[1]{\href{#1}{#1}}
\providecommand{\dodoi}[1]{doi:~\href{http://doi.org/#1}{\nolinkurl{#1}}}
\providecommand{\doeprint}[1]{\href{http://ascl.net/#1}{\nolinkurl{http://ascl.net/#1}}}
\providecommand{\doarXiv}[1]{\href{https://arxiv.org/abs/#1}{\nolinkurl{https://arxiv.org/abs/#1}}}

\bibitem[{E. {Anders} \& N. {Grevesse}(1989){Anders} \& {Grevesse}}]{anders89}
{Anders}, E., \& {Grevesse}, N. 1989, \bibinfo{title}{{Abundances of the
  elements: Meteoritic and solar},} \gca, 53, 197,
  \dodoi{10.1016/0016-7037(89)90286-X}

\bibitem[{M.~E. {Anderson} \& R. {Sunyaev}(2018){Anderson} \&
  {Sunyaev}}]{anderson18}
{Anderson}, M.~E., \& {Sunyaev}, R. 2018, \bibinfo{title}{{FUV line emission,
  gas kinematics, and discovery of [Fe XXI] {\ensuremath{\lambda}}1354.1 in the
  sightline toward a filament in M87},} \aap, 617, A123,
  \dodoi{10.1051/0004-6361/201732510}

\bibitem[{K.~A. {Arnaud}(1996){Arnaud}}]{arnaud96}
{Arnaud}, K.~A. 1996, \bibinfo{title}{{XSPEC: The First Ten Years},} in
  Astronomical Society of the Pacific Conference Series, Vol. 101, Astronomical
  Data Analysis Software and Systems V, ed. G.~H. {Jacoby} \& J.~{Barnes}, 17

\bibitem[{ {Astropy Collaboration} {et~al.}(2013){Astropy Collaboration},
  {Robitaille}, {Tollerud}, {Greenfield}, {Droettboom}, {Bray}, {Aldcroft},
  {Davis}, {Ginsburg}, {Price-Whelan}, {Kerzendorf}, {Conley}, {Crighton},
  {Barbary}, {Muna}, {Ferguson}, {Grollier}, {Parikh}, {Nair}, {Unther},
  {Deil}, {Woillez}, {Conseil}, {Kramer}, {Turner}, {Singer}, {Fox}, {Weaver},
  {Zabalza}, {Edwards}, {Azalee Bostroem}, {Burke}, {Casey}, {Crawford},
  {Dencheva}, {Ely}, {Jenness}, {Labrie}, {Lim}, {Pierfederici}, {Pontzen},
  {Ptak}, {Refsdal}, {Servillat}, \& {Streicher}}]{astropy13}
{Astropy Collaboration}, {Robitaille}, T.~P., {Tollerud}, E.~J., {et~al.} 2013,
  \bibinfo{title}{{Astropy: A community Python package for astronomy},} \aap,
  558, A33, \dodoi{10.1051/0004-6361/201322068}

\bibitem[{ {Astropy Collaboration} {et~al.}(2018){Astropy Collaboration},
  {Price-Whelan}, {Sip{\H{o}}cz}, {G{\"u}nther}, {Lim}, {Crawford}, {Conseil},
  {Shupe}, {Craig}, {Dencheva}, {Ginsburg}, {VanderPlas}, {Bradley},
  {P{\'e}rez-Su{\'a}rez}, {de Val-Borro}, {Aldcroft}, {Cruz}, {Robitaille},
  {Tollerud}, {Ardelean}, {Babej}, {Bach}, {Bachetti}, {Bakanov}, {Bamford},
  {Barentsen}, {Barmby}, {Baumbach}, {Berry}, {Biscani}, {Boquien}, {Bostroem},
  {Bouma}, {Brammer}, {Bray}, {Breytenbach}, {Buddelmeijer}, {Burke},
  {Calderone}, {Cano Rodr{\'\i}guez}, {Cara}, {Cardoso}, {Cheedella}, {Copin},
  {Corrales}, {Crichton}, {D'Avella}, {Deil}, {Depagne}, {Dietrich}, {Donath},
  {Droettboom}, {Earl}, {Erben}, {Fabbro}, {Ferreira}, {Finethy}, {Fox},
  {Garrison}, {Gibbons}, {Goldstein}, {Gommers}, {Greco}, {Greenfield},
  {Groener}, {Grollier}, {Hagen}, {Hirst}, {Homeier}, {Horton}, {Hosseinzadeh},
  {Hu}, {Hunkeler}, {Ivezi{\'c}}, {Jain}, {Jenness}, {Kanarek}, {Kendrew},
  {Kern}, {Kerzendorf}, {Khvalko}, {King}, {Kirkby}, {Kulkarni}, {Kumar},
  {Lee}, {Lenz}, {Littlefair}, {Ma}, {Macleod}, {Mastropietro}, {McCully},
  {Montagnac}, {Morris}, {Mueller}, {Mumford}, {Muna}, {Murphy}, {Nelson},
  {Nguyen}, {Ninan}, {N{\"o}the}, {Ogaz}, {Oh}, {Parejko}, {Parley}, {Pascual},
  {Patil}, {Patil}, {Plunkett}, {Prochaska}, {Rastogi}, {Reddy Janga},
  {Sabater}, {Sakurikar}, {Seifert}, {Sherbert}, {Sherwood-Taylor}, {Shih},
  {Sick}, {Silbiger}, {Singanamalla}, {Singer}, {Sladen}, {Sooley},
  {Sornarajah}, {Streicher}, {Teuben}, {Thomas}, {Tremblay}, {Turner},
  {Terr{\'o}n}, {van Kerkwijk}, {de la Vega}, {Watkins}, {Weaver}, {Whitmore},
  {Woillez}, {Zabalza}, \& {Astropy Contributors}}]{astropy18}
{Astropy Collaboration}, {Price-Whelan}, A.~M., {Sip{\H{o}}cz}, B.~M., {et~al.}
  2018, \bibinfo{title}{{The Astropy Project: Building an Open-science Project
  and Status of the v2.0 Core Package},} \aj, 156, 123,
  \dodoi{10.3847/1538-3881/aabc4f}

\bibitem[{ {Astropy Collaboration} {et~al.}(2022){Astropy Collaboration},
  {Price-Whelan}, {Lim}, {Earl}, {Starkman}, {Bradley}, {Shupe}, {Patil},
  {Corrales}, {Brasseur}, {N{\"o}the}, {Donath}, {Tollerud}, {Morris},
  {Ginsburg}, {Vaher}, {Weaver}, {Tocknell}, {Jamieson}, {van Kerkwijk},
  {Robitaille}, {Merry}, {Bachetti}, {G{\"u}nther}, {Aldcroft},
  {Alvarado-Montes}, {Archibald}, {B{\'o}di}, {Bapat}, {Barentsen},
  {Baz{\'a}n}, {Biswas}, {Boquien}, {Burke}, {Cara}, {Cara}, {Conroy},
  {Conseil}, {Craig}, {Cross}, {Cruz}, {D'Eugenio}, {Dencheva}, {Devillepoix},
  {Dietrich}, {Eigenbrot}, {Erben}, {Ferreira}, {Foreman-Mackey}, {Fox},
  {Freij}, {Garg}, {Geda}, {Glattly}, {Gondhalekar}, {Gordon}, {Grant},
  {Greenfield}, {Groener}, {Guest}, {Gurovich}, {Handberg}, {Hart},
  {Hatfield-Dodds}, {Homeier}, {Hosseinzadeh}, {Jenness}, {Jones}, {Joseph},
  {Kalmbach}, {Karamehmetoglu}, {Ka{\l}uszy{\'n}ski}, {Kelley}, {Kern},
  {Kerzendorf}, {Koch}, {Kulumani}, {Lee}, {Ly}, {Ma}, {MacBride}, {Maljaars},
  {Muna}, {Murphy}, {Norman}, {O'Steen}, {Oman}, {Pacifici}, {Pascual},
  {Pascual-Granado}, {Patil}, {Perren}, {Pickering}, {Rastogi}, {Roulston},
  {Ryan}, {Rykoff}, {Sabater}, {Sakurikar}, {Salgado}, {Sanghi}, {Saunders},
  {Savchenko}, {Schwardt}, {Seifert-Eckert}, {Shih}, {Jain}, {Shukla}, {Sick},
  {Simpson}, {Singanamalla}, {Singer}, {Singhal}, {Sinha}, {Sip{\H{o}}cz},
  {Spitler}, {Stansby}, {Streicher}, {{\v{S}}umak}, {Swinbank}, {Taranu},
  {Tewary}, {Tremblay}, {de Val-Borro}, {Van Kooten}, {Vasovi{\'c}}, {Verma},
  {de Miranda Cardoso}, {Williams}, {Wilson}, {Winkel}, {Wood-Vasey}, {Xue},
  {Yoachim}, {Zhang}, {Zonca}, \& {Astropy Project Contributors}}]{astropy22}
{Astropy Collaboration}, {Price-Whelan}, A.~M., {Lim}, P.~L., {et~al.} 2022,
  \bibinfo{title}{{The Astropy Project: Sustaining and Growing a
  Community-oriented Open-source Project and the Latest Major Release (v5.0) of
  the Core Package},} \apj, 935, 167, \dodoi{10.3847/1538-4357/ac7c74}

\bibitem[{P.~N. {Best} {et~al.}(2005){Best}, {Kauffmann}, {Heckman},
  {Brinchmann}, {Charlot}, {Ivezi{\'c}}, \& {White}}]{best05}
{Best}, P.~N., {Kauffmann}, G., {Heckman}, T.~M., {et~al.} 2005,
  \bibinfo{title}{{The host galaxies of radio-loud active galactic nuclei: mass
  dependences, gas cooling and active galactic nuclei feedback},} \mnras, 362,
  25, \dodoi{10.1111/j.1365-2966.2005.09192.x}

\bibitem[{L. {B{\^i}rzan} {et~al.}(2004){B{\^i}rzan}, {Rafferty}, {McNamara},
  {Wise}, \& {Nulsen}}]{birzan04}
{B{\^i}rzan}, L., {Rafferty}, D.~A., {McNamara}, B.~R., {Wise}, M.~W., \&
  {Nulsen}, P.~E.~J. 2004, \bibinfo{title}{{A Systematic Study of Radio-induced
  X-Ray Cavities in Clusters, Groups, and Galaxies},} \apj, 607, 800,
  \dodoi{10.1086/383519}

\bibitem[{L. {B{\^\i}rzan} {et~al.}(2012){B{\^\i}rzan}, {Rafferty}, {Nulsen},
  {McNamara}, {R{\"o}ttgering}, {Wise}, \& {Mittal}}]{birzan12}
{B{\^\i}rzan}, L., {Rafferty}, D.~A., {Nulsen}, P.~E.~J., {et~al.} 2012,
  \bibinfo{title}{{The duty cycle of radio-mode feedback in complete samples of
  clusters},} \mnras, 427, 3468, \dodoi{10.1111/j.1365-2966.2012.22083.x}

\bibitem[{A.~D. {Bolatto} {et~al.}(2013){Bolatto}, {Wolfire}, \&
  {Leroy}}]{bolatto13}
{Bolatto}, A.~D., {Wolfire}, M., \& {Leroy}, A.~K. 2013, \bibinfo{title}{{The
  CO-to-H$_{2}$ Conversion Factor},} \araa, 51, 207,
  \dodoi{10.1146/annurev-astro-082812-140944}

\bibitem[{M.~S. {Calzadilla} {et~al.}(2019){Calzadilla}, {Russell}, {McDonald},
  {Fabian}, {Baum}, {Combes}, {Donahue}, {Edge}, {McNamara}, \&
  {Nulsen}}]{calzadilla19}
{Calzadilla}, M.~S., {Russell}, H.~R., {McDonald}, M.~A., {et~al.} 2019,
  \bibinfo{title}{{Revealing a Highly Dynamic Cluster Core in Abell 1664 with
  Chandra},} \apj, 875, 65, \dodoi{10.3847/1538-4357/ab09f6}

\bibitem[{V. {Cammelli} {et~al.}(2026){Cammelli}, {Gaspari}, {Piana},
  {Barbani}, {Stel}, {Brustio}, {Olivares}, {Salvestrini}, {Danehkar}, {Reefe},
  {Temi}, {Maccagni}, {Tombesi}, \& {Fournier}}]{cammelli26}
{Cammelli}, V., {Gaspari}, M., {Piana}, O., {et~al.} 2026,
  \bibinfo{title}{{BlackHoleWeather -- Jet-regulated chaotic cold accretion
  across the meso scale: Morphology and thermodynamics},} arXiv e-prints,
  arXiv:2605.27503, \dodoi{10.48550/arXiv.2605.27503}

\bibitem[{K.~W. {Cavagnolo} {et~al.}(2008){Cavagnolo}, {Donahue}, {Voit}, \&
  {Sun}}]{cavagnolo08}
{Cavagnolo}, K.~W., {Donahue}, M., {Voit}, G.~M., \& {Sun}, M. 2008,
  \bibinfo{title}{{An Entropy Threshold for Strong H{\ensuremath{\alpha}} and
  Radio Emission in the Cores of Galaxy Clusters},} \apjl, 683, L107,
  \dodoi{10.1086/591665}

\bibitem[{E. {Churazov} {et~al.}(2001){Churazov}, {Br{\"u}ggen}, {Kaiser},
  {B{\"o}hringer}, \& {Forman}}]{churazov01}
{Churazov}, E., {Br{\"u}ggen}, M., {Kaiser}, C.~R., {B{\"o}hringer}, H., \&
  {Forman}, W. 2001, \bibinfo{title}{{Evolution of Buoyant Bubbles in M87},}
  \apj, 554, 261, \dodoi{10.1086/321357}

\bibitem[{E. {Churazov} {et~al.}(2000){Churazov}, {Forman}, {Jones}, \&
  {B{\"o}hringer}}]{churazov00}
{Churazov}, E., {Forman}, W., {Jones}, C., \& {B{\"o}hringer}, H. 2000,
  \bibinfo{title}{{Asymmetric, arc minute scale structures around NGC 1275},}
  \aap, 356, 788, \dodoi{10.48550/arXiv.astro-ph/0002375}

\bibitem[{C. {Cicone} {et~al.}(2014){Cicone}, {Maiolino}, {Sturm},
  {Graci{\'a}-Carpio}, {Feruglio}, {Neri}, {Aalto}, {Davies}, {Fiore},
  {Fischer}, {Garc{\'\i}a-Burillo}, {Gonz{\'a}lez-Alfonso}, {Hailey-Dunsheath},
  {Piconcelli}, \& {Veilleux}}]{cicone14}
{Cicone}, C., {Maiolino}, R., {Sturm}, E., {et~al.} 2014,
  \bibinfo{title}{{Massive molecular outflows and evidence for AGN feedback
  from CO observations},} \aap, 562, A21, \dodoi{10.1051/0004-6361/201322464}

\bibitem[{D.~A. {Dale} {et~al.}(2014){Dale}, {Helou}, {Magdis}, {Armus},
  {D{\'\i}az-Santos}, \& {Shi}}]{dale14}
{Dale}, D.~A., {Helou}, G., {Magdis}, G.~E., {et~al.} 2014, \bibinfo{title}{{A
  Two-parameter Model for the Infrared/Submillimeter/Radio Spectral Energy
  Distributions of Galaxies and Active Galactic Nuclei},} \apj, 784, 83,
  \dodoi{10.1088/0004-637X/784/1/83}

\bibitem[{S. {Diehl} {et~al.}(2008){Diehl}, {Li}, {Fryer}, \&
  {Rafferty}}]{diehl08}
{Diehl}, S., {Li}, H., {Fryer}, C.~L., \& {Rafferty}, D. 2008,
  \bibinfo{title}{{Constraining the Nature of X-Ray Cavities in Clusters and
  Galaxies},} \apj, 687, 173, \dodoi{10.1086/591310}

\bibitem[{M. {Donahue} {et~al.}(2011){Donahue}, {de Messi{\`e}res},
  {O'Connell}, {Voit}, {Hoffer}, {McNamara}, \& {Nulsen}}]{donahue11}
{Donahue}, M., {de Messi{\`e}res}, G.~E., {O'Connell}, R.~W., {et~al.} 2011,
  \bibinfo{title}{{Polycyclic Aromatic Hydrocarbons, Ionized Gas, and Molecular
  Hydrogen in Brightest Cluster Galaxies of Cool-core Clusters of Galaxies},}
  \apj, 732, 40, \dodoi{10.1088/0004-637X/732/1/40}

\bibitem[{M. {Donahue} \& G.~M. {Voit}(2022){Donahue} \& {Voit}}]{donahue22}
{Donahue}, M., \& {Voit}, G.~M. 2022, \bibinfo{title}{{Baryon cycles in the
  biggest galaxies},} \physrep, 973, 1, \dodoi{10.1016/j.physrep.2022.04.005}

\bibitem[{X. {Duan} \& F. {Guo}(2024){Duan} \& {Guo}}]{duan24}
{Duan}, X., \& {Guo}, F. 2024, \bibinfo{title}{{Cold Filaments Formed in Hot
  Wake Flows Uplifted by Active Galactic Nucleus Bubbles in Galaxy Clusters},}
  \apj, 972, 41, \dodoi{10.3847/1538-4357/ad5bdf}

\bibitem[{A.~C. {Edge}(2001){Edge}}]{edge01}
{Edge}, A.~C. 2001, \bibinfo{title}{{The detection of molecular gas in the
  central galaxies of cooling flow clusters},} \mnras, 328, 762,
  \dodoi{10.1046/j.1365-8711.2001.04802.x}

\bibitem[{A.~C. {Edge} \& D.~T. {Frayer}(2003){Edge} \& {Frayer}}]{edge03}
{Edge}, A.~C., \& {Frayer}, D.~T. 2003, \bibinfo{title}{{Resolving Molecular
  gas in the Central Galaxies of Cooling Flow Clusters},} \apjl, 594, L13,
  \dodoi{10.1086/378386}

\bibitem[{A.~C. {Edge} {et~al.}(2010){Edge}, {Oonk}, {Mittal}, {Allen}, {Baum},
  {B{\"o}hringer}, {Bregman}, {Bremer}, {Combes}, {Crawford}, {Donahue},
  {Egami}, {Fabian}, {Ferland}, {Hamer}, {Hatch}, {Jaffe}, {Johnstone},
  {McNamara}, {O'Dea}, {Popesso}, {Quillen}, {Salom{\'e}}, {Sarazin}, {Voit},
  {Wilman}, \& {Wise}}]{edge10}
{Edge}, A.~C., {Oonk}, J.~B.~R., {Mittal}, R., {et~al.} 2010,
  \bibinfo{title}{{Herschel photometry of brightest cluster galaxies in cooling
  flow clusters},} \aap, 518, L47, \dodoi{10.1051/0004-6361/201014572}

\bibitem[{A.~C. {Fabian}(2012){Fabian}}]{fabian12}
{Fabian}, A.~C. 2012, \bibinfo{title}{{Observational Evidence of Active
  Galactic Nuclei Feedback},} \araa, 50, 455,
  \dodoi{10.1146/annurev-astro-081811-125521}

\bibitem[{A.~C. {Fabian} {et~al.}(2002){Fabian}, {Allen}, {Crawford},
  {Johnstone}, {Morris}, {Sanders}, \& {Schmidt}}]{fabian02}
{Fabian}, A.~C., {Allen}, S.~W., {Crawford}, C.~S., {et~al.} 2002,
  \bibinfo{title}{{The missing soft X-ray luminosity in cluster cooling
  flows},} \mnras, 332, L50, \dodoi{10.1046/j.1365-8711.2002.05510.x}

\bibitem[{A.~C. {Fabian} {et~al.}(2008){Fabian}, {Johnstone}, {Sanders},
  {Conselice}, {Crawford}, {Gallagher}, \& {Zweibel}}]{fabian08}
{Fabian}, A.~C., {Johnstone}, R.~M., {Sanders}, J.~S., {et~al.} 2008,
  \bibinfo{title}{{Magnetic support of the optical emission line filaments in
  NGC 1275},} \nat, 454, 968, \dodoi{10.1038/nature07169}

\bibitem[{A.~C. {Fabian} {et~al.}(2023{\natexlab{a}}){Fabian}, {Sanders},
  {Ferland}, {McNamara}, {Pinto}, \& {Walker}}]{fabian23}
{Fabian}, A.~C., {Sanders}, J.~S., {Ferland}, G.~J., {et~al.}
  2023{\natexlab{a}}, \bibinfo{title}{{Hidden Cooling Flows in clusters of
  Galaxies II: a wider sample},} \mnras, 521, 1794,
  \dodoi{10.1093/mnras/stad507}

\bibitem[{A.~C. {Fabian} {et~al.}(2023{\natexlab{b}}){Fabian}, {Sanders},
  {Ferland}, {McNamara}, {Pinto}, \& {Walker}}]{fabian23b}
{Fabian}, A.~C., {Sanders}, J.~S., {Ferland}, G.~J., {et~al.}
  2023{\natexlab{b}}, \bibinfo{title}{{Hidden cooling flows in clusters of
  galaxies - III. Accretion on to the central black hole},} \mnras, 524, 716,
  \dodoi{10.1093/mnras/stad1870}

\bibitem[{A.~C. {Fabian} {et~al.}(2025){Fabian}, {Sanders}, {Ferland},
  {Russell}, {McNamara}, {Pinto}, \& {Walker}}]{fabian25}
{Fabian}, A.~C., {Sanders}, J.~S., {Ferland}, G.~J., {et~al.} 2025,
  \bibinfo{title}{{Hidden (absorbed) Cooling Flows V: Groups and Galaxies
  including Spirals},} arXiv e-prints, arXiv:2508.14785,
  \dodoi{10.48550/arXiv.2508.14785}

\bibitem[{A.~C. {Fabian} {et~al.}(2011){Fabian}, {Sanders}, {Williams},
  {Lazarian}, {Ferland}, \& {Johnstone}}]{fabian11}
{Fabian}, A.~C., {Sanders}, J.~S., {Williams}, R.~J.~R., {et~al.} 2011,
  \bibinfo{title}{{The energy source of the filaments around the giant galaxy
  NGC 1275},} \mnras, 417, 172, \dodoi{10.1111/j.1365-2966.2011.19034.x}

\bibitem[{A.~C. {Fabian} {et~al.}(2016){Fabian}, {Walker}, {Russell}, {Pinto},
  {Canning}, {Salome}, {Sanders}, {Taylor}, {Zweibel}, {Conselice}, {Combes},
  {Crawford}, {Ferland}, {Gallagher}, {Hatch}, {Johnstone}, \&
  {Reynolds}}]{fabian16}
{Fabian}, A.~C., {Walker}, S.~A., {Russell}, H.~R., {et~al.} 2016,
  \bibinfo{title}{{HST imaging of the dusty filaments and nucleus swirl in
  NGC4696 at the centre of the Centaurus Cluster},} \mnras, 461, 922,
  \dodoi{10.1093/mnras/stw1350}

\bibitem[{R.~J. {Farber} \& M. {Gronke}(2023){Farber} \& {Gronke}}]{farber23}
{Farber}, R.~J., \& {Gronke}, M. 2023, \bibinfo{title}{{Multiphase
  fragmentation: Molecular shattering},} \mnras, \dodoi{10.1093/mnras/stad2373}

\bibitem[{C. {Feruglio} {et~al.}(2010){Feruglio}, {Maiolino}, {Piconcelli},
  {Menci}, {Aussel}, {Lamastra}, \& {Fiore}}]{feruglio10}
{Feruglio}, C., {Maiolino}, R., {Piconcelli}, E., {et~al.} 2010,
  \bibinfo{title}{{Quasar feedback revealed by giant molecular outflows},}
  \aap, 518, L155, \dodoi{10.1051/0004-6361/201015164}

\bibitem[{D.~B. {Fielding} \& G.~L. {Bryan}(2022){Fielding} \&
  {Bryan}}]{fielding22}
{Fielding}, D.~B., \& {Bryan}, G.~L. 2022, \bibinfo{title}{{The Structure of
  Multiphase Galactic Winds},} \apj, 924, 82, \dodoi{10.3847/1538-4357/ac2f41}

\bibitem[{F. {Fiore} {et~al.}(2017){Fiore}, {Feruglio}, {Shankar}, {Bischetti},
  {Bongiorno}, {Brusa}, {Carniani}, {Cicone}, {Duras}, {Lamastra}, {Mainieri},
  {Marconi}, {Menci}, {Maiolino}, {Piconcelli}, {Vietri}, \&
  {Zappacosta}}]{fiore17}
{Fiore}, F., {Feruglio}, C., {Shankar}, F., {et~al.} 2017, \bibinfo{title}{{AGN
  wind scaling relations and the co-evolution of black holes and galaxies},}
  \aap, 601, A143, \dodoi{10.1051/0004-6361/201629478}

\bibitem[{A. {Fluetsch} {et~al.}(2019){Fluetsch}, {Maiolino}, {Carniani},
  {Marconi}, {Cicone}, {Bourne}, {Costa}, {Fabian}, {Ishibashi}, \&
  {Venturi}}]{fluetsch19}
{Fluetsch}, A., {Maiolino}, R., {Carniani}, S., {et~al.} 2019,
  \bibinfo{title}{{Cold molecular outflows in the local Universe and their
  feedback effect on galaxies},} \mnras, 483, 4586,
  \dodoi{10.1093/mnras/sty3449}

\bibitem[{M. {Fournier} {et~al.}(2026){Fournier}, {Grete}, {Br{\"u}ggen},
  {O'Shea}, {Voit}, {Wibking}, \& {Prasad}}]{fournier26}
{Fournier}, M., {Grete}, P., {Br{\"u}ggen}, M., {et~al.} 2026,
  \bibinfo{title}{{XMAGNET -- Stir before serving: a Lagrangian perspective on
  mixing-driven condensation in the intracluster medium},} arXiv e-prints,
  arXiv:2605.00563.
\newblock \doarXiv{2605.00563}

\bibitem[{P.~E. {Freeman} {et~al.}(2002){Freeman}, {Kashyap}, {Rosner}, \&
  {Lamb}}]{freeman02}
{Freeman}, P.~E., {Kashyap}, V., {Rosner}, R., \& {Lamb}, D.~Q. 2002,
  \bibinfo{title}{{A Wavelet-Based Algorithm for the Spatial Analysis of
  Poisson Data},} \apjs, 138, 185, \dodoi{10.1086/324017}

\bibitem[{M. {Gaspari} {et~al.}(2017){Gaspari}, {Temi}, \&
  {Brighenti}}]{gaspari17}
{Gaspari}, M., {Temi}, P., \& {Brighenti}, F. 2017, \bibinfo{title}{{Raining on
  black holes and massive galaxies: the top-down multiphase condensation
  model},} \mnras, 466, 677, \dodoi{10.1093/mnras/stw3108}

\bibitem[{M. {Gaspari} {et~al.}(2018){Gaspari}, {McDonald}, {Hamer},
  {Brighenti}, {Temi}, {Gendron-Marsolais}, {Hlavacek-Larrondo}, {Edge},
  {Werner}, {Tozzi}, {Sun}, {Stone}, {Tremblay}, {Hogan}, {Eckert}, {Ettori},
  {Yu}, {Biffi}, \& {Planelles}}]{gaspari18}
{Gaspari}, M., {McDonald}, M., {Hamer}, S.~L., {et~al.} 2018,
  \bibinfo{title}{{Shaken Snow Globes: Kinematic Tracers of the Multiphase
  Condensation Cascade in Massive Galaxies, Groups, and Clusters},} \apj, 854,
  167, \dodoi{10.3847/1538-4357/aaaa1b}

\bibitem[{F. {Gastaldello} {et~al.}(2009){Gastaldello}, {Buote}, {Temi},
  {Brighenti}, {Mathews}, \& {Ettori}}]{gastaldello09}
{Gastaldello}, F., {Buote}, D.~A., {Temi}, P., {et~al.} 2009,
  \bibinfo{title}{{X-Ray Cavities, Filaments, and Cold Fronts in the Core of
  the Galaxy Group NGC 5044},} \apj, 693, 43,
  \dodoi{10.1088/0004-637X/693/1/43}

\bibitem[{M.-J. {Gingras} {et~al.}(2025){Gingras}, {McNamara}, {Coil},
  {Perrotta}, {Brighenti}, {Russell}, {Oh}, \& {Ning}}]{gingras25}
{Gingras}, M.-J., {McNamara}, B.~R., {Coil}, A.~L., {et~al.} 2025,
  \bibinfo{title}{{Star Formation Histories and Stellar Dynamics in the Central
  Galaxies of RX J0820.9+0752, A1835, and PKS 0745-191},} arXiv e-prints,
  arXiv:2511.01986, \dodoi{10.48550/arXiv.2511.01986}

\bibitem[{M.-J. {Gingras} {et~al.}(2024){Gingras}, {Coil}, {McNamara},
  {Perrotta}, {Brighenti}, {Russell}, {Li}, {Oh}, \& {Ning}}]{gingras24}
{Gingras}, M.-J., {Coil}, A.~L., {McNamara}, B.~R., {et~al.} 2024,
  \bibinfo{title}{{Complex Kinematics of Nebular Gas in Active Galaxies
  Centered in Cooling X-Ray Atmospheres},} \apj, 977, 159,
  \dodoi{10.3847/1538-4357/ad822a}

\bibitem[{S.~L. {Hamer} {et~al.}(2016){Hamer}, {Edge}, {Swinbank}, {Wilman},
  {Combes}, {Salom{\'e}}, {Fabian}, {Crawford}, {Russell}, {Hlavacek-Larrondo},
  {McNamara}, \& {Bremer}}]{hamer16}
{Hamer}, S.~L., {Edge}, A.~C., {Swinbank}, A.~M., {et~al.} 2016,
  \bibinfo{title}{{Optical emission line nebulae in galaxy cluster cores 1: the
  morphological, kinematic and spectral properties of the sample},} \mnras,
  460, 1758, \dodoi{10.1093/mnras/stw1054}

\bibitem[{ {Hitomi Collaboration} {et~al.}(2016){Hitomi Collaboration},
  {Aharonian}, {Akamatsu}, {Akimoto}, {Allen}, {Anabuki}, {Angelini}, {Arnaud},
  {Audard}, {Awaki}, {Axelsson}, {Bamba}, {Bautz}, {Blandford}, {Brenneman},
  {Brown}, {Bulbul}, {Cackett}, {Chernyakova}, {Chiao}, {Coppi}, {Costantini},
  {de Plaa}, {den Herder}, {Done}, {Dotani}, {Ebisawa}, {Eckart}, {Enoto},
  {Ezoe}, {Fabian}, {Ferrigno}, {Foster}, {Fujimoto}, {Fukazawa}, {Furuzawa},
  {Galeazzi}, {Gallo}, {Gandhi}, {Giustini}, {Goldwurm}, {Gu}, {Guainazzi},
  {Haba}, {Hagino}, {Hamaguchi}, {Harrus}, {Hatsukade}, {Hayashi}, {Hayashi},
  {Hayashida}, {Hiraga}, {Hornschemeier}, {Hoshino}, {Hughes}, {Iizuka},
  {Inoue}, {Inoue}, {Ishibashi}, {Ishida}, {Ishikawa}, {Ishisaki}, {Itoh},
  {Iyomoto}, {Kaastra}, {Kallman}, {Kamae}, {Kara}, {Kataoka}, {Katsuda},
  {Katsuta}, {Kawaharada}, {Kawai}, {Kelley}, {Khangulyan}, {Kilbourne},
  {King}, {Kitaguchi}, {Kitamoto}, {Kitayama}, {Kohmura}, {Kokubun}, {Koyama},
  {Koyama}, {Kretschmar}, {Krimm}, {Kubota}, {Kunieda}, {Laurent}, {Lebrun},
  {Lee}, {Leutenegger}, {Limousin}, {Loewenstein}, {Long}, {Lumb}, {Madejski},
  {Maeda}, {Maier}, {Makishima}, {Markevitch}, {Matsumoto}, {Matsushita},
  {McCammon}, {McNamara}, {Mehdipour}, {Miller}, {Miller}, {Mineshige},
  {Mitsuda}, {Mitsuishi}, {Miyazawa}, {Mizuno}, {Mori}, {Mori}, {Moseley},
  {Mukai}, {Murakami}, {Murakami}, {Mushotzky}, {Nagino}, {Nakagawa},
  {Nakajima}, {Nakamori}, {Nakano}, {Nakashima}, {Nakazawa}, {Nobukawa},
  {Noda}, {Nomachi}, {O'Dell}, {Odaka}, {Ohashi}, {Ohno}, {Okajima}, {Ota},
  {Ozaki}, {Paerels}, {Paltani}, {Parmar}, {Petre}, {Pinto}, {Pohl}, {Porter},
  {Pottschmidt}, {Ramsey}, {Reynolds}, {Russell}, {Safi-Harb}, {Saito},
  {Sakai}, {Sameshima}, {Sato}, {Sato}, {Sato}, {Sawada}, {Schartel},
  {Serlemitsos}, {Seta}, {Shidatsu}, {Simionescu}, {Smith}, {Soong}, {Stawarz},
  {Sugawara}, {Sugita}, {Szymkowiak}, {Tajima}, {Takahashi}, {Takahashi},
  {Takeda}, {Takei}, {Tamagawa}, {Tamura}, {Tamura}, {Tanaka}, {Tanaka},
  {Tanaka}, {Tashiro}, {Tawara}, {Terada}, {Terashima}, {Tombesi}, {Tomida},
  {Tsuboi}, {Tsujimoto}, {Tsunemi}, {Tsuru}, {Uchida}, {Uchiyama}, {Uchiyama},
  {Ueda}, {Ueda}, {Ueno}, {Uno}, {Urry}, {Ursino}, {de Vries}, {Watanabe},
  {Werner}, {Wik}, {Wilkins}, {Williams}, {Yamada}, {Yamaguchi}, {Yamaoka},
  {Yamasaki}, {Yamauchi}, {Yamauchi}, {Yaqoob}, {Yatsu}, {Yonetoku}, {Yoshida},
  {Yuasa}, {Zhuravleva}, \& {Zoghbi}}]{hitomi16}
{Hitomi Collaboration}, {Aharonian}, F., {Akamatsu}, H., {et~al.} 2016,
  \bibinfo{title}{{The quiescent intracluster medium in the core of the Perseus
  cluster},} \nat, 535, 117, \dodoi{10.1038/nature18627}

\bibitem[{ {Hitomi Collaboration} {et~al.}(2018){Hitomi Collaboration},
  {Aharonian}, {Akamatsu}, {Akimoto}, {Allen}, {Angelini}, {Audard}, {Awaki},
  {Axelsson}, {Bamba}, {Bautz}, {Blandford}, {Brenneman}, {Brown}, {Bulbul},
  {Cackett}, {Canning}, {Chernyakova}, {Chiao}, {Coppi}, {Costantini}, {de
  Plaa}, {de Vries}, {den Herder}, {Done}, {Dotani}, {Ebisawa}, {Eckart},
  {Enoto}, {Ezoe}, {Fabian}, {Ferrigno}, {Foster}, {Fujimoto}, {Fukazawa},
  {Furuzawa}, {Galeazzi}, {Gallo}, {Gandhi}, {Giustini}, {Goldwurm}, {Gu},
  {Guainazzi}, {Haba}, {Hagino}, {Hamaguchi}, {Harrus}, {Hatsukade}, {Hayashi},
  {Hayashi}, {Hayashi}, {Hayashida}, {Hiraga}, {Hornschemeier}, {Hoshino},
  {Hughes}, {Ichinohe}, {Iizuka}, {Inoue}, {Inoue}, {Inoue}, {Ishida},
  {Ishikawa}, {Ishisaki}, {Iwai}, {Kaastra}, {Kallman}, {Kamae}, {Kataoka},
  {Katsuda}, {Kawai}, {Kelley}, {Kilbourne}, {Kitaguchi}, {Kitamoto},
  {Kitayama}, {Kohmura}, {Kokubun}, {Koyama}, {Koyama}, {Kretschmar}, {Krimm},
  {Kubota}, {Kunieda}, {Laurent}, {Lee}, {Leutenegger}, {Limousin},
  {Loewenstein}, {Long}, {Lumb}, {Madejski}, {Maeda}, {Maier}, {Makishima},
  {Markevitch}, {Matsumoto}, {Matsushita}, {McCammon}, {McNamara}, {Mehdipour},
  {Miller}, {Miller}, {Mineshige}, {Mitsuda}, {Mitsuishi}, {Miyazawa},
  {Mizuno}, {Mori}, {Mori}, {Mukai}, {Murakami}, {Mushotzky}, {Nakagawa},
  {Nakajima}, {Nakamori}, {Nakashima}, {Nakazawa}, {Nobukawa}, {Nobukawa},
  {Noda}, {Odaka}, {Ohashi}, {Ohno}, {Okajima}, {Ota}, {Ozaki}, {Paerels},
  {Paltani}, {Petre}, {Pinto}, {Porter}, {Pottschmidt}, {Reynolds},
  {Safi-Harb}, {Saito}, {Sakai}, {Sasaki}, {Sato}, {Sato}, {Sato}, {Sawada},
  {Schartel}, {Serlemtsos}, {Seta}, {Shidatsu}, {Simionescu}, {Smith}, {Soong},
  {Stawarz}, {Sugawara}, {Sugita}, {Szymkowiak}, {Tajima}, {Takahashi},
  {Takahashi}, {Takeda}, {Takei}, {Tamagawa}, {Tamura}, {Tanaka}, {Tanaka},
  {Tanaka}, {Tanaka}, {Tashiro}, {Tawara}, {Terada}, {Terashima}, {Tombesi},
  {Tomida}, {Tsuboi}, {Tsujimoto}, {Tsunemi}, {Tsuru}, {Uchida}, {Uchiyama},
  {Uchiyama}, {Ueda}, {Ueda}, {Uno}, {Urry}, {Ursino}, {Wang}, {Watanabe},
  {Werner}, {Wilkins}, {Williams}, {Yamada}, {Yamaguchi}, {Yamaoka},
  {Yamasaki}, {Yamauchi}, {Yamauchi}, {Yaqoob}, {Yatsu}, {Yonetoku},
  {Zhuravleva}, \& {Zoghbi}}]{hitomi18}
{Hitomi Collaboration}, {Aharonian}, F., {Akamatsu}, H., {et~al.} 2018,
  \bibinfo{title}{{Atmospheric gas dynamics in the Perseus cluster observed
  with Hitomi},} \pasj, 70, 9, \dodoi{10.1093/pasj/psx138}

\bibitem[{M.~T. {Hogan} {et~al.}(2017{\natexlab{a}}){Hogan}, {McNamara},
  {Pulido}, {Nulsen}, {Russell}, {Vantyghem}, {Edge}, \& {Main}}]{hogan17a}
{Hogan}, M.~T., {McNamara}, B.~R., {Pulido}, F., {et~al.} 2017{\natexlab{a}},
  \bibinfo{title}{{Mass Distribution in Galaxy Cluster Cores},} \apj, 837, 51,
  \dodoi{10.3847/1538-4357/aa5f56}

\bibitem[{M.~T. {Hogan} {et~al.}(2017{\natexlab{b}}){Hogan}, {McNamara},
  {Pulido}, {Nulsen}, {Vantyghem}, {Russell}, {Edge}, {Babyk}, {Main}, \&
  {McDonald}}]{hogan17b}
{Hogan}, M.~T., {McNamara}, B.~R., {Pulido}, F.~A., {et~al.}
  2017{\natexlab{b}}, \bibinfo{title}{{The Onset of Thermally Unstable Cooling
  from the Hot Atmospheres of Giant Galaxies in Clusters: Constraints on
  Feedback Models},} \apj, 851, 66, \dodoi{10.3847/1538-4357/aa9af3}

\bibitem[{F. {Jennings} {et~al.}(2023){Jennings}, {Beckmann}, {Sijacki}, \&
  {Dubois}}]{jennings23}
{Jennings}, F., {Beckmann}, R.~S., {Sijacki}, D., \& {Dubois}, Y. 2023,
  \bibinfo{title}{{Shattering and growth of cold clouds in galaxy clusters: the
  role of radiative cooling, magnetic fields, and thermal conduction},} \mnras,
  518, 5215, \dodoi{10.1093/mnras/stac3426}

\bibitem[{P.~M.~W. {Kalberla} {et~al.}(2005){Kalberla}, {Burton}, {Hartmann},
  {Arnal}, {Bajaja}, {Morras}, \& {P{\"o}ppel}}]{kalberla05}
{Kalberla}, P.~M.~W., {Burton}, W.~B., {Hartmann}, D., {et~al.} 2005,
  \bibinfo{title}{{The Leiden/Argentine/Bonn (LAB) Survey of Galactic HI. Final
  data release of the combined LDS and IAR surveys with improved
  stray-radiation corrections},} \aap, 440, 775,
  \dodoi{10.1051/0004-6361:20041864}

\bibitem[{B.~C. {Kelly}(2007){Kelly}}]{kelly07}
{Kelly}, B.~C. 2007, \bibinfo{title}{{Some Aspects of Measurement Error in
  Linear Regression of Astronomical Data},} \apj, 665, 1489,
  \dodoi{10.1086/519947}

\bibitem[{C.~C. {Kirkpatrick} \& B.~R. {McNamara}(2015){Kirkpatrick} \&
  {McNamara}}]{kirkpatrick15}
{Kirkpatrick}, C.~C., \& {McNamara}, B.~R. 2015, \bibinfo{title}{{Hot outflows
  in galaxy clusters},} \mnras, 452, 4361, \dodoi{10.1093/mnras/stv1574}

\bibitem[{G. {Kokotanekov} {et~al.}(2018){Kokotanekov}, {Wise}, {de Vries}, \&
  {Intema}}]{kokotanekov18}
{Kokotanekov}, G., {Wise}, M.~W., {de Vries}, M., \& {Intema}, H.~T. 2018,
  \bibinfo{title}{{Signatures of multiple episodes of AGN activity in the core
  of Abell 1795},} \aap, 618, A152, \dodoi{10.1051/0004-6361/201833222}

\bibitem[{Y. {Li} \& G.~L. {Bryan}(2014{\natexlab{a}}){Li} \& {Bryan}}]{li14}
{Li}, Y., \& {Bryan}, G.~L. 2014{\natexlab{a}}, \bibinfo{title}{{Modeling
  Active Galactic Nucleus Feedback in Cool-core Clusters: The Formation of Cold
  Clumps},} \apj, 789, 153, \dodoi{10.1088/0004-637X/789/2/153}

\bibitem[{Y. {Li} \& G.~L. {Bryan}(2014{\natexlab{b}}){Li} \& {Bryan}}]{li14a}
{Li}, Y., \& {Bryan}, G.~L. 2014{\natexlab{b}}, \bibinfo{title}{{Modeling
  Active Galactic Nucleus Feedback in Cool-core Clusters: The Formation of Cold
  Clumps},} \apj, 789, 153, \dodoi{10.1088/0004-637X/789/2/153}

\bibitem[{C.~G. {Martz} {et~al.}(2020){Martz}, {McNamara}, {Nulsen},
  {Vantyghem}, {Gingras}, {Babyk}, {Russell}, {Edge}, {McDonald}, {Tamhane},
  {Fabian}, \& {Hogan}}]{martz20}
{Martz}, C.~G., {McNamara}, B.~R., {Nulsen}, P.~E.~J., {et~al.} 2020,
  \bibinfo{title}{{Thermally Unstable Cooling Stimulated by Uplift: The Spoiler
  Clusters},} \apj, 897, 57, \dodoi{10.3847/1538-4357/ab96cd}

\bibitem[{W.~G. {Mathews} \& F. {Brighenti}(2003){Mathews} \&
  {Brighenti}}]{mathews03}
{Mathews}, W.~G., \& {Brighenti}, F. 2003, \bibinfo{title}{{Hot Gas in and
  around Elliptical Galaxies},} \araa, 41, 191,
  \dodoi{10.1146/annurev.astro.41.090401.094542}

\bibitem[{P. {Mazzotta} \& S. {Giacintucci}(2008){Mazzotta} \&
  {Giacintucci}}]{mazzotta08}
{Mazzotta}, P., \& {Giacintucci}, S. 2008, \bibinfo{title}{{Do Radio Core-Halos
  and Cold Fronts in Non-Major-Merging Clusters Originate from the Same Gas
  Sloshing?},} \apjl, 675, L9, \dodoi{10.1086/529433}

\bibitem[{P. {Mazzotta} {et~al.}(2001){Mazzotta}, {Markevitch}, {Forman},
  {Jones}, {Vikhlinin}, \& {VanSpeybroeck}}]{mazzotta01}
{Mazzotta}, P., {Markevitch}, M., {Forman}, W.~R., {et~al.} 2001,
  \bibinfo{title}{{Chandra Observation of MS 1455.0+2232: cold fronts in a
  massive cooling flow cluster?},} arXiv e-prints, astro,
  \dodoi{10.48550/arXiv.astro-ph/0108476}

\bibitem[{M. {McCourt} {et~al.}(2018){McCourt}, {Oh}, {O'Leary}, \&
  {Madigan}}]{mccourt18}
{McCourt}, M., {Oh}, S.~P., {O'Leary}, R., \& {Madigan}, A.-M. 2018,
  \bibinfo{title}{{A characteristic scale for cold gas},} \mnras, 473, 5407,
  \dodoi{10.1093/mnras/stx2687}

\bibitem[{M. {McCourt} {et~al.}(2012){McCourt}, {Sharma}, {Quataert}, \&
  {Parrish}}]{mccourt12}
{McCourt}, M., {Sharma}, P., {Quataert}, E., \& {Parrish}, I.~J. 2012,
  \bibinfo{title}{{Thermal instability in gravitationally stratified plasmas:
  implications for multiphase structure in clusters and galaxy haloes},}
  \mnras, 419, 3319, \dodoi{10.1111/j.1365-2966.2011.19972.x}

\bibitem[{M. {McDonald} {et~al.}(2018){McDonald}, {Gaspari}, {McNamara}, \&
  {Tremblay}}]{mcdonald18}
{McDonald}, M., {Gaspari}, M., {McNamara}, B.~R., \& {Tremblay}, G.~R. 2018,
  \bibinfo{title}{{Revisiting the Cooling Flow Problem in Galaxies, Groups, and
  Clusters of Galaxies},} \apj, 858, 45, \dodoi{10.3847/1538-4357/aabace}

\bibitem[{M. {McDonald} {et~al.}(2010){McDonald}, {Veilleux}, {Rupke}, \&
  {Mushotzky}}]{mcdonald10}
{McDonald}, M., {Veilleux}, S., {Rupke}, D. S.~N., \& {Mushotzky}, R. 2010,
  \bibinfo{title}{{On the Origin of the Extended H{\ensuremath{\alpha}}
  Filaments in Cooling Flow Clusters},} \apj, 721, 1262,
  \dodoi{10.1088/0004-637X/721/2/1262}

\bibitem[{M. {McDonald} {et~al.}(2014){McDonald}, {Swinbank}, {Edge}, {Wilner},
  {Veilleux}, {Benson}, {Hogan}, {Marrone}, {McNamara}, {Wei}, {Bayliss}, \&
  {Bautz}}]{mcdonald14b}
{McDonald}, M., {Swinbank}, M., {Edge}, A.~C., {et~al.} 2014,
  \bibinfo{title}{{The State of the Warm and Cold Gas in the Extreme Starburst
  at the Core of the Phoenix Galaxy Cluster (SPT-CLJ2344-4243)},} \apj, 784,
  18, \dodoi{10.1088/0004-637X/784/1/18}

\bibitem[{M. {McDonald} {et~al.}(2019){McDonald}, {McNamara}, {Voit},
  {Bayliss}, {Benson}, {Brodwin}, {Canning}, {Florian}, {Garmire}, {Gaspari},
  {Gladders}, {Hlavacek-Larrondo}, {Kara}, {Reichardt}, {Russell}, {Saro},
  {Sharon}, {Somboonpanyakul}, {Tremblay}, \& {van Weeren}}]{mcdonald19}
{McDonald}, M., {McNamara}, B.~R., {Voit}, G.~M., {et~al.} 2019,
  \bibinfo{title}{{Anatomy of a Cooling Flow: The Feedback Response to Pure
  Cooling in the Core of the Phoenix Cluster},} \apj, 885, 63,
  \dodoi{10.3847/1538-4357/ab464c}

\bibitem[{B.~R. {McNamara} {et~al.}(2026){McNamara}, {Fabian}, {Russell},
  {Nulsen}, {Simionescu}, {Majumder}, {Miller}, \& {Sarkar}}]{mcnamara26}
{McNamara}, B.~R., {Fabian}, A.~C., {Russell}, H.~R., {et~al.} 2026,
  \bibinfo{title}{{Are X-ray Atmospheres Heated by Turbulent Dissipation? XRISM
  Constraints},} arXiv e-prints, arXiv:2604.19607,
  \dodoi{10.48550/arXiv.2604.19607}

\bibitem[{B.~R. {McNamara} \& P.~E.~J. {Nulsen}(2007){McNamara} \&
  {Nulsen}}]{mcnamara07}
{McNamara}, B.~R., \& {Nulsen}, P.~E.~J. 2007, \bibinfo{title}{{Heating Hot
  Atmospheres with Active Galactic Nuclei},} \araa, 45, 117,
  \dodoi{10.1146/annurev.astro.45.051806.110625}

\bibitem[{B.~R. {McNamara} \& P.~E.~J. {Nulsen}(2012){McNamara} \&
  {Nulsen}}]{mcnamara12}
{McNamara}, B.~R., \& {Nulsen}, P.~E.~J. 2012, \bibinfo{title}{{Mechanical
  feedback from active galactic nuclei in galaxies, groups and clusters},} New
  Journal of Physics, 14, 055023, \dodoi{10.1088/1367-2630/14/5/055023}

\bibitem[{B.~R. {McNamara} {et~al.}(2016){McNamara}, {Russell}, {Nulsen},
  {Hogan}, {Fabian}, {Pulido}, \& {Edge}}]{mcnamara16}
{McNamara}, B.~R., {Russell}, H.~R., {Nulsen}, P.~E.~J., {et~al.} 2016,
  \bibinfo{title}{{A Mechanism for Stimulating AGN Feedback by Lifting Gas in
  Massive Galaxies},} \apj, 830, 79, \dodoi{10.3847/0004-637X/830/2/79}

\bibitem[{B.~R. {McNamara} {et~al.}(2006){McNamara}, {Rafferty}, {B{\^\i}rzan},
  {Steiner}, {Wise}, {Nulsen}, {Carilli}, {Ryan}, \& {Sharma}}]{mcnamara06}
{McNamara}, B.~R., {Rafferty}, D.~A., {B{\^\i}rzan}, L., {et~al.} 2006,
  \bibinfo{title}{{The Starburst in the Abell 1835 Cluster Central Galaxy: A
  Case Study of Galaxy Formation Regulated by an Outburst from a Supermassive
  Black Hole},} \apj, 648, 164, \dodoi{10.1086/505859}

\bibitem[{B.~R. {McNamara} {et~al.}(2014){McNamara}, {Russell}, {Nulsen},
  {Edge}, {Murray}, {Main}, {Vantyghem}, {Combes}, {Fabian}, {Salome},
  {Kirkpatrick}, {Baum}, {Bregman}, {Donahue}, {Egami}, {Hamer}, {O'Dea},
  {Oonk}, {Tremblay}, \& {Voit}}]{mcnamara14}
{McNamara}, B.~R., {Russell}, H.~R., {Nulsen}, P.~E.~J., {et~al.} 2014,
  \bibinfo{title}{{A 10$^{10}$ Solar Mass Flow of Molecular Gas in the A1835
  Brightest Cluster Galaxy},} \apj, 785, 44, \dodoi{10.1088/0004-637X/785/1/44}

\bibitem[{R. {Mohapatra} {et~al.}(2023){Mohapatra}, {Sharma}, {Federrath}, \&
  {Quataert}}]{mohapatra23}
{Mohapatra}, R., {Sharma}, P., {Federrath}, C., \& {Quataert}, E. 2023,
  \bibinfo{title}{{Multiphase condensation in cluster haloes: interplay of
  cooling, buoyancy, and mixing},} \mnras, 525, 3831,
  \dodoi{10.1093/mnras/stad2574}

\bibitem[{P.~E.~J. {Nulsen} {et~al.}(2005){Nulsen}, {McNamara}, {Wise}, \&
  {David}}]{nulsen05}
{Nulsen}, P.~E.~J., {McNamara}, B.~R., {Wise}, M.~W., \& {David}, L.~P. 2005,
  \bibinfo{title}{{The Cluster-Scale AGN Outburst in Hydra A},} \apj, 628, 629,
  \dodoi{10.1086/430845}

\bibitem[{V. {Olivares} {et~al.}(2025){Olivares}, {Picquenot}, {Su}, {Gaspari},
  {Gendron-Marsolais}, {Polles}, \& {Nulsen}}]{olivares25}
{Olivares}, V., {Picquenot}, A., {Su}, Y., {et~al.} 2025, \bibinfo{title}{{An
  H{\ensuremath{\alpha}}-X-ray surface-brightness correlation for filaments in
  cooling-flow clusters},} Nature Astronomy, 9, 449,
  \dodoi{10.1038/s41550-024-02473-8}

\bibitem[{V. {Olivares} {et~al.}(2019){Olivares}, {Salome}, {Combes}, {Hamer},
  {Guillard}, {Lehnert}, {Polles}, {Beckmann}, {Dubois}, {Donahue}, {Edge},
  {Fabian}, {McNamara}, {Rose}, {Russell}, {Tremblay}, {Vantyghem}, {Canning},
  {Ferland }, {Godard}, {Peirani}, \& {Pineau des Forets}}]{olivares19}
{Olivares}, V., {Salome}, P., {Combes}, F., {et~al.} 2019,
  \bibinfo{title}{{Ubiquitous cold and massive filaments in cool core
  clusters},} \aap, 631, A22, \dodoi{10.1051/0004-6361/201935350}

\bibitem[{V. {Olivares} {et~al.}(2023){Olivares}, {Su}, {Forman}, {Gaspari},
  {Andrade-Santos}, {Salome}, {Nulsen}, {Edge}, {Combes}, \&
  {Jones}}]{olivares23}
{Olivares}, V., {Su}, Y., {Forman}, W., {et~al.} 2023, \bibinfo{title}{{X-Ray
  Cavity Dynamics and Their Role in the Gas Precipitation in Planck
  Sunyaev-Zeldovich (SZ) Selected Clusters},} \apj, 954, 56,
  \dodoi{10.3847/1538-4357/ace359}

\bibitem[{M.~B. {Pandge} {et~al.}(2019){Pandge}, {Sonkamble}, {Parekh},
  {Dabhade}, {Parmar}, {Patil}, \& {Raychaudhury}}]{pandge19}
{Pandge}, M.~B., {Sonkamble}, S.~S., {Parekh}, V., {et~al.} 2019,
  \bibinfo{title}{{AGN Feedback in Galaxy Groups: A Detailed Study of X-Ray
  Features and Diffuse Radio Emission in IC 1262},} \apj, 870, 62,
  \dodoi{10.3847/1538-4357/aaf105}

\bibitem[{R. {Paterno-Mahler} {et~al.}(2013){Paterno-Mahler}, {Blanton},
  {Randall}, \& {Clarke}}]{pm13}
{Paterno-Mahler}, R., {Blanton}, E.~L., {Randall}, S.~W., \& {Clarke}, T.~E.
  2013, \bibinfo{title}{{Deep Chandra Observations of the Extended Gas Sloshing
  Spiral in A2029},} \apj, 773, 114, \dodoi{10.1088/0004-637X/773/2/114}

\bibitem[{J.~R. {Peterson} {et~al.}(2001){Peterson}, {Paerels}, {Kaastra},
  {Arnaud}, {Reiprich}, {Fabian}, {Mushotzky}, {Jernigan}, \&
  {Sakelliou}}]{peterson01}
{Peterson}, J.~R., {Paerels}, F.~B.~S., {Kaastra}, J.~S., {et~al.} 2001,
  \bibinfo{title}{{X-ray imaging-spectroscopy of Abell 1835},} \aap, 365, L104,
  \dodoi{10.1051/0004-6361:20000021}

\bibitem[{F. {Pizzolato} \& N. {Soker}(2005){Pizzolato} \&
  {Soker}}]{pizzolato05}
{Pizzolato}, F., \& {Soker}, N. 2005, \bibinfo{title}{{On the Nature of
  Feedback Heating in Cooling Flow Clusters},} \apj, 632, 821,
  \dodoi{10.1086/444344}

\bibitem[{E.~C.~D. {Pope} {et~al.}(2010){Pope}, {Babul}, {Pavlovski}, {Bower},
  \& {Dotter}}]{pope10}
{Pope}, E. C.~D., {Babul}, A., {Pavlovski}, G., {Bower}, R.~G., \& {Dotter}, A.
  2010, \bibinfo{title}{{Mass transport by buoyant bubbles in galaxy
  clusters},} \mnras, 406, 2023, \dodoi{10.1111/j.1365-2966.2010.16816.x}

\bibitem[{D. {Prasad} {et~al.}(2015){Prasad}, {Sharma}, \& {Babul}}]{prasad15}
{Prasad}, D., {Sharma}, P., \& {Babul}, A. 2015, \bibinfo{title}{{Cool Core
  Cycles: Cold Gas and AGN Jet Feedback in Cluster Cores},} \apj, 811, 108,
  \dodoi{10.1088/0004-637X/811/2/108}

\bibitem[{D. {Prasad} {et~al.}(2022){Prasad}, {Voit}, \& {O'Shea}}]{prasad22}
{Prasad}, D., {Voit}, G.~M., \& {O'Shea}, B.~W. 2022,
  \bibinfo{title}{{Atmospheric Circulation in Simulations of the AGN-CGM
  Connection at Halo Masses 10$^{13.5}$ M $_{{\ensuremath{\odot}}}$},} \apj,
  932, 18, \dodoi{10.3847/1538-4357/ac69ee}

\bibitem[{F.~A. {Pulido} {et~al.}(2018){Pulido}, {McNamara}, {Edge}, {Hogan},
  {Vantyghem}, {Russell}, {Nulsen}, {Babyk}, \& {Salom{\'e}}}]{pulido18}
{Pulido}, F.~A., {McNamara}, B.~R., {Edge}, A.~C., {et~al.} 2018,
  \bibinfo{title}{{The Origin of Molecular Clouds in Central Galaxies},} \apj,
  853, 177, \dodoi{10.3847/1538-4357/aaa54b}

\bibitem[{D.~A. {Rafferty} {et~al.}(2006){Rafferty}, {McNamara}, {Nulsen}, \&
  {Wise}}]{rafferty06}
{Rafferty}, D.~A., {McNamara}, B.~R., {Nulsen}, P.~E.~J., \& {Wise}, M.~W.
  2006, \bibinfo{title}{{The Feedback-regulated Growth of Black Holes and
  Bulges through Gas Accretion and Starbursts in Cluster Central Dominant
  Galaxies},} \apj, 652, 216, \dodoi{10.1086/507672}

\bibitem[{M. {Reefe} {et~al.}(2025){Reefe}, {McDonald}, {Chatzikos}, {Seebeck},
  {Mushotzky}, {Veilleux}, {Allen}, {Bayliss}, {Calzadilla}, {Canning},
  {Floyd}, {Gaspari}, {Hlavacek-Larrondo}, {McNamara}, {Russell}, {Sharon}, \&
  {Somboonpanyakul}}]{reefe25a}
{Reefe}, M., {McDonald}, M., {Chatzikos}, M., {et~al.} 2025,
  \bibinfo{title}{{Directly imaging the cooling flow in the Phoenix cluster},}
  \nat, 638, 360, \dodoi{10.1038/s41586-024-08369-x}

\bibitem[{Y. {Revaz} {et~al.}(2008){Revaz}, {Combes}, \&
  {Salom{\'e}}}]{revaz08}
{Revaz}, Y., {Combes}, F., \& {Salom{\'e}}, P. 2008, \bibinfo{title}{{Formation
  of cold filaments in cooling flow clusters},} \aap, 477, L33,
  \dodoi{10.1051/0004-6361:20078915}

\bibitem[{C.~J. {Riseley} {et~al.}(2022){Riseley}, {Rajpurohit}, {Loi},
  {Botteon}, {Timmerman}, {Biava}, {Bonafede}, {Bonnassieux}, {Brunetti},
  {En{\ss}lin}, {Di Gennaro}, {Ignesti}, {Shimwell}, {Stuardi}, {Vernstrom}, \&
  {van Weeren}}]{riseley22}
{Riseley}, C.~J., {Rajpurohit}, K., {Loi}, F., {et~al.} 2022,
  \bibinfo{title}{{A MeerKAT-meets-LOFAR study of MS 1455.0 + 2232: a 590
  kiloparsec 'mini'-halo in a sloshing cool-core cluster},} \mnras, 512, 4210,
  \dodoi{10.1093/mnras/stac672}

\bibitem[{H.~R. {Russell} {et~al.}(2026){Russell}, {Nulsen}, {Fabian},
  {McNamara}, {Sanders}, \& {Werner}}]{russell26}
{Russell}, H.~R., {Nulsen}, P.~E.~J., {Fabian}, A.~C., {et~al.} 2026,
  \bibinfo{title}{{The splash beneath the largest radio bubble in a cluster
  core},} arXiv e-prints, arXiv:2604.14292, \dodoi{10.48550/arXiv.2604.14292}

\bibitem[{H.~R. {Russell} {et~al.}(2008){Russell}, {Sanders}, \&
  {Fabian}}]{russell08}
{Russell}, H.~R., {Sanders}, J.~S., \& {Fabian}, A.~C. 2008,
  \bibinfo{title}{{Direct X-ray spectral deprojection of galaxy clusters},}
  \mnras, 390, 1207, \dodoi{10.1111/j.1365-2966.2008.13823.x}

\bibitem[{H.~R. {Russell} {et~al.}(2017){Russell}, {McNamara}, {Fabian},
  {Nulsen}, {Combes}, {Edge}, {Hogan}, {McDonald}, {Salom{\'e}}, {Tremblay}, \&
  {Vantyghem}}]{russell17b}
{Russell}, H.~R., {McNamara}, B.~R., {Fabian}, A.~C., {et~al.} 2017,
  \bibinfo{title}{{Close entrainment of massive molecular gas flows by radio
  bubbles in the central galaxy of Abell 1795},} \mnras, 472, 4024,
  \dodoi{10.1093/mnras/stx2255}

\bibitem[{H.~R. {Russell} {et~al.}(2019){Russell}, {McNamara}, {Fabian},
  {Nulsen}, {Combes}, {Edge}, {Madar}, {Olivares}, {Salom{\'e}}, \&
  {Vantyghem}}]{russell19}
{Russell}, H.~R., {McNamara}, B.~R., {Fabian}, A.~C., {et~al.} 2019,
  \bibinfo{title}{{Driving massive molecular gas flows in central cluster
  galaxies with AGN feedback},} \mnras, 490, 3025,
  \dodoi{10.1093/mnras/stz2719}

\bibitem[{P. {Salom{\'e}} \& F. {Combes}(2003){Salom{\'e}} \&
  {Combes}}]{salome03}
{Salom{\'e}}, P., \& {Combes}, F. 2003, \bibinfo{title}{{Cold molecular gas in
  cooling flow clusters of galaxies},} \aap, 412, 657,
  \dodoi{10.1051/0004-6361:20031438}

\bibitem[{J.~S. {Sanders} \& A.~C. {Fabian}(2007){Sanders} \&
  {Fabian}}]{sanders07}
{Sanders}, J.~S., \& {Fabian}, A.~C. 2007, \bibinfo{title}{{A deeper X-ray
  study of the core of the Perseus galaxy cluster: the power of sound waves and
  the distribution of metals and cosmic rays},} \mnras, 381, 1381,
  \dodoi{10.1111/j.1365-2966.2007.12347.x}

\bibitem[{J. {Shin} {et~al.}(2016){Shin}, {Woo}, \& {Mulchaey}}]{shin16}
{Shin}, J., {Woo}, J.-H., \& {Mulchaey}, J.~S. 2016, \bibinfo{title}{{A
  Systematic Search for X-Ray Cavities in Galaxy Clusters, Groups, and
  Elliptical Galaxies},} \apjs, 227, 31, \dodoi{10.3847/1538-4365/227/2/31}

\bibitem[{J. {Silk} \& J.~R. {Burke}(1974){Silk} \& {Burke}}]{silk74}
{Silk}, J., \& {Burke}, J.~R. 1974, \bibinfo{title}{{Dust Grains in a Hot Gas.
  11. Astrophysical Applications},} \apj, 190, 11, \dodoi{10.1086/152841}

\bibitem[{A. {Simionescu} {et~al.}(2009){Simionescu}, {Werner},
  {B{\"o}hringer}, {Kaastra}, {Finoguenov}, {Br{\"u}ggen}, \&
  {Nulsen}}]{simionescu09}
{Simionescu}, A., {Werner}, N., {B{\"o}hringer}, H., {et~al.} 2009,
  \bibinfo{title}{{Chemical enrichment in the cluster of galaxies Hydra A},}
  \aap, 493, 409, \dodoi{10.1051/0004-6361:200810225}

\bibitem[{A. {Simionescu} {et~al.}(2008){Simionescu}, {Werner}, {Finoguenov},
  {B{\"o}hringer}, \& {Br{\"u}ggen}}]{simionescu08}
{Simionescu}, A., {Werner}, N., {Finoguenov}, A., {B{\"o}hringer}, H., \&
  {Br{\"u}ggen}, M. 2008, \bibinfo{title}{{Metal-rich multi-phase gas in M 87.
  AGN-driven metal transport, magnetic-field supported multi-temperature gas,
  and constraints on non-thermal emission observed with XMM-Newton},} \aap,
  482, 97, \dodoi{10.1051/0004-6361:20078749}

\bibitem[{N. {Soker} {et~al.}(2002){Soker}, {Blanton}, \& {Sarazin}}]{soker02}
{Soker}, N., {Blanton}, E.~L., \& {Sarazin}, C.~L. 2002, \bibinfo{title}{{Hot
  Bubbles in Cooling Flow Clusters},} \apj, 573, 533, \dodoi{10.1086/340799}

\bibitem[{N. {Soker} {et~al.}(2004){Soker}, {Blanton}, \& {Sarazin}}]{soker04}
{Soker}, N., {Blanton}, E.~L., \& {Sarazin}, C.~L. 2004,
  \bibinfo{title}{{Cooling of X-ray emitting gas by heat conduction in the
  center of cooling flow clusters},} \aap, 422, 445,
  \dodoi{10.1051/0004-6361:20034415}

\bibitem[{S. {Sotira} {et~al.}(2026){Sotira}, {Bourne}, {Sijacki}, {Vazza}, \&
  {Brighenti}}]{sotira26}
{Sotira}, S., {Bourne}, M.~A., {Sijacki}, D., {Vazza}, F., \& {Brighenti}, F.
  2026, \bibinfo{title}{{Cold gas formation triggered by active galactic nuclei
  jet feedback in galaxy cluster cores},} arXiv e-prints, arXiv:2601.14391,
  \dodoi{10.48550/arXiv.2601.14391}

\bibitem[{P. {Tamhane} {et~al.}(2026){Tamhane}, {Sun}, {Waldron}, {Hosogi}, {da
  Silva}, {Le}, {Gaspari}, {Combes}, {Werner}, {Schellenberger}, {Fabian},
  {Canning}, {David}, {Donahue}, \& {Voit}}]{tamhane26}
{Tamhane}, P., {Sun}, M., {Waldron}, W., {et~al.} 2026, \bibinfo{title}{{HST
  view of NGC 5044: Constraints on filament widths, magnetic support,
  multiphase structure, and comparison with cluster environments},} \pasa, 43,
  e036, \dodoi{10.1017/pasa.2026.10172}

\bibitem[{P.~D. {Tamhane} {et~al.}(2022){Tamhane}, {McNamara}, {Russell},
  {Edge}, {Fabian}, {Nulsen}, \& {Babyk}}]{tamhane22}
{Tamhane}, P.~D., {McNamara}, B.~R., {Russell}, H.~R., {et~al.} 2022,
  \bibinfo{title}{{Molecular flows in contemporary active galaxies and the
  efficacy of radio-mechanical feedback},} \mnras, 516, 861,
  \dodoi{10.1093/mnras/stac2168}

\bibitem[{P.~D. {Tamhane} {et~al.}(2023){Tamhane}, {McNamara}, {Russell},
  {Combes}, {Qiu}, {Edge}, {Maiolino}, {Fabian}, {Nulsen}, {Johnstone}, \&
  {Carniani}}]{tamhane23}
{Tamhane}, P.~D., {McNamara}, B.~R., {Russell}, H.~R., {et~al.} 2023,
  \bibinfo{title}{{Radio jet-ISM interaction and positive radio-mechanical
  feedback in Abell 1795},} \mnras, 519, 3338, \dodoi{10.1093/mnras/stac3803}

\bibitem[{T. {Tamura} {et~al.}(2001){Tamura}, {Bleeker}, {Kaastra}, {Ferrigno},
  \& {Molendi}}]{tamura01a}
{Tamura}, T., {Bleeker}, J.~A.~M., {Kaastra}, J.~S., {Ferrigno}, C., \&
  {Molendi}, S. 2001, \bibinfo{title}{{XMM-Newton observations of the cluster
  of galaxies Abell 496. Measurements of the elemental abundances in the
  intracluster medium},} \aap, 379, 107, \dodoi{10.1051/0004-6361:20011317}

\bibitem[{B. {Tan} \& D.~B. {Fielding}(2024){Tan} \& {Fielding}}]{tan24}
{Tan}, B., \& {Fielding}, D.~B. 2024, \bibinfo{title}{{Cloud atlas: navigating
  the multiphase landscape of tempestuous galactic winds},} \mnras, 527, 9683,
  \dodoi{10.1093/mnras/stad3793}

\bibitem[{P. {Temi} {et~al.}(2026){Temi}, {Ubertosi}, {Brighenti},
  {Maragkoudakis}, {Olivares}, {Amblard}, {Gaspari}, {Gitti}, {Marcum},
  {Fogarty}, {Borlaff}, \& {Mathews}}]{temi26}
{Temi}, P., {Ubertosi}, F., {Brighenti}, F., {et~al.} 2026,
  \bibinfo{title}{{Active Galactic Nucleus Feedback and the Development of
  Dusty Multiphase Gas in X-Ray Emitting Elliptical Galaxies},} \apj, 1000,
  144, \dodoi{10.3847/1538-4357/ae4972}

\bibitem[{M. {Valentini} \& F. {Brighenti}(2015){Valentini} \&
  {Brighenti}}]{valentini15}
{Valentini}, M., \& {Brighenti}, F. 2015, \bibinfo{title}{{AGN-stimulated
  cooling of hot gas in elliptical galaxies},} \mnras, 448, 1979,
  \dodoi{10.1093/mnras/stv090}

\bibitem[{A.~N. {Vantyghem} {et~al.}(2014){Vantyghem}, {McNamara}, {Russell},
  {Main}, {Nulsen}, {Wise}, {Hoekstra}, \& {Gitti}}]{vantyghem14}
{Vantyghem}, A.~N., {McNamara}, B.~R., {Russell}, H.~R., {et~al.} 2014,
  \bibinfo{title}{{Cycling of the powerful AGN in MS 0735.6+7421 and the duty
  cycle of radio AGN in clusters},} \mnras, 442, 3192,
  \dodoi{10.1093/mnras/stu1030}

\bibitem[{A.~N. {Vantyghem} {et~al.}(2018){Vantyghem}, {McNamara}, {Russell},
  {Edge}, {Nulsen}, {Combes}, {Fabian}, {McDonald}, \&
  {Salom{\'e}}}]{vantyghem18}
{Vantyghem}, A.~N., {McNamara}, B.~R., {Russell}, H.~R., {et~al.} 2018,
  \bibinfo{title}{{Molecular Gas Filaments and Star-forming Knots Beneath an
  X-Ray Cavity in RXC J1504-0248},} \apj, 863, 193,
  \dodoi{10.3847/1538-4357/aad2e0}

\bibitem[{A.~N. {Vantyghem} {et~al.}(2021){Vantyghem}, {McNamara}, {O'Dea},
  {Baum}, {Combes}, {Edge}, {Fabian}, {McDonald}, {Nulsen}, {Russell}, \&
  {Salom{\'e}}}]{vantyghem21}
{Vantyghem}, A.~N., {McNamara}, B.~R., {O'Dea}, C.~P., {et~al.} 2021,
  \bibinfo{title}{{A Massive, Clumpy Molecular Gas Distribution and Displaced
  AGN in Zw 3146},} \apj, 910, 53, \dodoi{10.3847/1538-4357/abe306}

\bibitem[{G.~M. {Voit} \& M. {Donahue}(2015){Voit} \& {Donahue}}]{voit15}
{Voit}, G.~M., \& {Donahue}, M. 2015, \bibinfo{title}{{Cooling Time, Freefall
  Time, and Precipitation in the Cores of ACCEPT Galaxy Clusters},} \apjl, 799,
  L1, \dodoi{10.1088/2041-8205/799/1/L1}

\bibitem[{G.~M. {Voit} {et~al.}(2017){Voit}, {Meece}, {Li}, {O'Shea}, {Bryan},
  \& {Donahue}}]{voit17}
{Voit}, G.~M., {Meece}, G., {Li}, Y., {et~al.} 2017, \bibinfo{title}{{A Global
  Model for Circumgalactic and Cluster-core Precipitation},} \apj, 845, 80,
  \dodoi{10.3847/1538-4357/aa7d04}

\bibitem[{G.~M. {Voit} {et~al.}(2026){Voit}, {Wibking}, \& {Yaldiz}}]{voit26}
{Voit}, G.~M., {Wibking}, B.~D., \& {Yaldiz}, D. 2026,
  \bibinfo{title}{{Magnetohydrodynamic Precipitation},} \pasp, 138, 073001,
  \dodoi{10.1088/1538-3873/ae7d55}

\bibitem[{N. {Werner} {et~al.}(2010){Werner}, {Simionescu}, {Million}, {Allen},
  {Nulsen}, {von der Linden}, {Hansen}, {B{\"o}hringer}, {Churazov}, {Fabian},
  {Forman}, {Jones}, {Sanders}, \& {Taylor}}]{werner10}
{Werner}, N., {Simionescu}, A., {Million}, E.~T., {et~al.} 2010,
  \bibinfo{title}{{Feedback under the microscope-II. Heating, gas uplift and
  mixing in the nearest cluster core},} \mnras, 407, 2063,
  \dodoi{10.1111/j.1365-2966.2010.16755.x}

\bibitem[{M.~W. {Wise} {et~al.}(2007){Wise}, {McNamara}, {Nulsen}, {Houck}, \&
  {David}}]{wise07}
{Wise}, M.~W., {McNamara}, B.~R., {Nulsen}, P.~E.~J., {Houck}, J.~C., \&
  {David}, L.~P. 2007, \bibinfo{title}{{X-Ray Supercavities in the Hydra A
  Cluster and the Outburst History of the Central Galaxy's Active Nucleus},}
  \apj, 659, 1153, \dodoi{10.1086/512767}

\bibitem[{ {Xrism Collaboration} {et~al.}(2025){Xrism Collaboration}, {Audard},
  {Awaki}, {Ballhausen}, {Bamba}, {Behar}, {Boissay-Malaquin}, {Brenneman},
  {Brown}, {Corrales}, {Costantini}, {Cumbee}, {Diaz Trigo}, {Done}, {Dotani},
  {Ebisawa}, {Eckart}, {Eckert}, {Eguchi}, {Enoto}, {Ezoe}, {Foster},
  {Fujimoto}, {Fujita}, {Fukazawa}, {Fukushima}, {Furuzawa}, {Gallo},
  {Garc{\'\i}a}, {Gu}, {Guainazzi}, {Hagino}, {Hamaguchi}, {Hatsukade},
  {Hayashi}, {Hayashi}, {Hell}, {Hodges-Kluck}, {Hornschemeier}, {Ichinohe},
  {Ishida}, {Ishikawa}, {Ishisaki}, {Kaastra}, {Kallman}, {Kara}, {Katsuda},
  {Kanemaru}, {Kelley}, {Kilbourne}, {Kitamoto}, {Kobayashi}, {Kohmura},
  {Kubota}, {Leutenegger}, {Loewenstein}, {Maeda}, {Markevitch}, {Matsumoto},
  {Matsushita}, {McCammon}, {McNamara}, {Mernier}, {Miller}, {Miller},
  {Mitsuishi}, {Mizumoto}, {Mizuno}, {Mori}, {Mukai}, {Murakami}, {Mushotzky},
  {Nakajima}, {Nakazawa}, {Ness}, {Nobukawa}, {Nobukawa}, {Noda}, {Odaka},
  {Ogawa}, {Ogorzalek}, {Okajima}, {Ota}, {Paltani}, {Petre}, {Plucinsky},
  {Porter}, {Pottschmidt}, {Sato}, {Sato}, {Sawada}, {Seta}, {Shidatsu},
  {Simionescu}, {Smith}, {Suzuki}, {Szymkowiak}, {Takahashi}, {Takeo},
  {Tamagawa}, {Tamura}, {Tanaka}, {Tanimoto}, {Tashiro}, {Terada}, {Terashima},
  {Tsuboi}, {Tsujimoto}, {Tsunemi}, {Tsuru}, {Uchida}, {Uchida}, {Uchida},
  {Uchiyama}, {Ueda}, {Uno}, {Vink}, {Watanabe}, {Williams}, {Yamada},
  {Yamada}, {Yamaguchi}, {Yamaoka}, {Yamasaki}, {Yamauchi}, {Yamauchi},
  {Yaqoob}, {Yoneyama}, {Yoshida}, {Yukita}, {Zhuravleva}, {Bartalesi},
  {Ettori}, {Kosarzycki}, {Lovisari}, {Rose}, {Sarkar}, {Sun}, \&
  {Tamhane}}]{xrism_a2029}
{Xrism Collaboration}, {Audard}, M., {Awaki}, H., {et~al.} 2025,
  \bibinfo{title}{{XRISM Reveals Low Nonthermal Pressure in the Core of the
  Hot, Relaxed Galaxy Cluster A2029},} \apjl, 982, L5,
  \dodoi{10.3847/2041-8213/ada7cd}

\bibitem[{ {XRISM Collaboration} {et~al.}(2025){XRISM Collaboration}, {Audard},
  {Awaki}, {Ballhausen}, {Bamba}, {Behar}, {Boissay-Malaquin}, {Brenneman},
  {Brown}, {Corrales}, {Costantini}, {Cumbee}, {Done}, {Dotani}, {Ebisawa},
  {Eckart}, {Eckert}, {Enoto}, {Eguchi}, {Ezoe}, {Foster}, {Fujimoto},
  {Fujita}, {Fukazawa}, {Fukushima}, {Furuzawa}, {Gallo}, {Garc{\'\i}a}, {Gu},
  {Guainazzi}, {Hagino}, {Hamaguchi}, {Hatsukade}, {Hayashi}, {Hayashi},
  {Hell}, {Hodges-Kluck}, {Hornschemeier}, {Ichinohe}, {Ishida}, {Ishikawa},
  {Ishisaki}, {Kaastra}, {Kallman}, {Kara}, {Katsuda}, {Kanemaru}, {Kelley},
  {Kilbourne}, {Kitamoto}, {Kobayashi}, {Kohmura}, {Kubota}, {Leutenegger},
  {Loewenstein}, {Maeda}, {Markevitch}, {Matsumoto}, {Matsushita}, {McCammon},
  {McNamara}, {Mernier}, {Miller}, {Miller}, {Mitsuishi}, {Mizumoto}, {Mizuno},
  {Mori}, {Mukai}, {Murakami}, {Mushotzky}, {Nakajima}, {Nakazawa}, {Ness},
  {Nobukawa}, {Nobukawa}, {Noda}, {Odaka}, {Ogawa}, {Ogorzalek}, {Okajima},
  {Ota}, {Paltani}, {Petre}, {Plucinsky}, {Porter}, {Pottschmidt}, {Sato},
  {Sato}, {Sawada}, {Seta}, {Shidatsu}, {Simionescu}, {Smith}, {Suzuki},
  {Szymkowiak}, {Takahashi}, {Takeo}, {Tamagawa}, {Tamura}, {Tanaka},
  {Tanimoto}, {Tashiro}, {Terada}, {Terashima}, {Trigo}, {Tsuboi}, {Tsujimoto},
  {Tsunemi}, {Tsuru}, {Uchida}, {Uchida}, {Uchida}, {Uchiyama}, {Ueda}, {Uno},
  {Vink}, {Watanabe}, {Williams}, {Yamada}, {Yamada}, {Yamaguchi}, {Yamaoka},
  {Yamasaki}, {Yamauchi}, {Yamauchi}, {Yaqoob}, {Yoneyama}, {Yoshida},
  {Yukita}, {Zhuravleva}, {Kondo}, {Werner}, {Pl{\v{s}}ek}, {Sun}, {Hosogi}, \&
  {Majumder}}]{xrism_centaurus}
{XRISM Collaboration}, {Audard}, M., {Awaki}, H., {et~al.} 2025,
  \bibinfo{title}{{The bulk motion of gas in the core of the Centaurus galaxy
  cluster},} \nat, 638, 365, \dodoi{10.1038/s41586-024-08561-z}

\bibitem[{C. {Zhang} {et~al.}(2018){Zhang}, {Churazov}, \&
  {Schekochihin}}]{zhang18}
{Zhang}, C., {Churazov}, E., \& {Schekochihin}, A.~A. 2018,
  \bibinfo{title}{{Generation of internal waves by buoyant bubbles in galaxy
  clusters and heating of intracluster medium},} \mnras, 478, 4785,
  \dodoi{10.1093/mnras/sty1269}

\bibitem[{C. {Zhang} {et~al.}(2022){Zhang}, {Zhuravleva}, {Gendron-Marsolais},
  {Churazov}, {Schekochihin}, \& {Forman}}]{zhang22}
{Zhang}, C., {Zhuravleva}, I., {Gendron-Marsolais}, M.-L., {et~al.} 2022,
  \bibinfo{title}{{Bubble-driven gas uplift in galaxy clusters and its velocity
  features},} \mnras, 517, 616, \dodoi{10.1093/mnras/stac2282}

\end{thebibliography}
\bibliographystyle{aasjournalv7}



\end{document}